\documentclass[aps,prl,twocolumn,superscriptaddress,10pt]{revtex4-1}
\usepackage[colorlinks,bookmarks=true,citecolor=blue,linkcolor=blue,urlcolor=blue]{hyperref}
\usepackage[dvipsnames]{xcolor}
\usepackage{amsmath,amssymb,graphicx}
\usepackage{newtxtext,newtxmath,mathrsfs,multirow,enumitem}
\usepackage[export]{adjustbox}
\usepackage[boxsize=.5em, aligntableaux=center]{ytableau}
\usepackage{cancel}
\usepackage{comment}

\setlist{align=parleft,leftmargin=0pt,itemindent=2.5\parindent,labelsep=0em,labelwidth=1.5\parindent,nosep}

\graphicspath{{./fig/}}

\usepackage{xcolor}
\definecolor{orange}{rgb}{1,0.5,0}

\begin{document}

\title{
Emergence of 3D Superconformal Ising Criticality on the Fuzzy Sphere 
}

\author{Yin Tang}
\affiliation{Westlake Institute of Advanced Study, Westlake University, Hangzhou 310024, China}
\affiliation{Institute for Advanced Study, Kyushu University, Fukuoka 819-0395, Japan}

\author{Cristian Voinea}
\affiliation{School of Physics and Astronomy, University of Leeds, Leeds LS2 9JT, United Kingdom}

\author{Liangdong Hu}
\affiliation{Westlake Institute of Advanced Study, Westlake University, Hangzhou 310024, China}

\author{Zlatko Papi\'c}
\affiliation{School of Physics and Astronomy, University of Leeds, Leeds LS2 9JT, United Kingdom}

\author{W. Zhu}
\affiliation{Westlake Institute of Advanced Study, Westlake University, Hangzhou 310024, China}

\begin{abstract}
Supersymmetric conformal field theories (SCFTs) form a unique subset of quantum field theories which provide powerful insights into strongly coupled critical phenomena. 
Here, we present a microscopic and non-perturbative realization of the three-dimensional 
$\mathcal{N}=1$ superconformal Ising critical point, based on a Yukawa-type coupling between a 3D Ising CFT and a gauged Majorana fermion. Using the recently developed fuzzy-sphere regularization, we directly extract the scaling dimensions of low-lying operators via the state–operator correspondence.
At the critical point, we demonstrate conformal multiplet structure together with the hallmark of emergent spacetime supersymmetry through characteristic relations between fermionic and bosonic operators. Moreover, by tuning the Yukawa coupling, we explicitly track the evolution of operator spectra from the decoupled Ising–Majorana fixed point to the interacting superconformal fixed point, revealing renormalization-group flow at the operator level.
Our results establish a controlled, non-perturbative microscopic route to 3D SCFTs.

\end{abstract}

\maketitle

{\sl Introduction.---}Supersymmetry (SUSY) posits a spacetime symmetry that interchanges bosonic and fermionic degrees of freedom, which imposes powerful constraints on quantum field theories \cite{Sohnius:1985qm,Wess_Bagger_Book,Weinberg:2000cr}. Separately, conformal field theory (CFT) provides an exact, non-perturbative framework for describing a wide range of scale-invariant critical phenomena in nature~\cite{Cardy_book,yellowbook}. 
When SUSY meets CFT, the resulting supersymmetric conformal field theories (SCFTs) exhibit rigid mathematical structures, such as highly constrained operator spectra, with deep connections to dualities and anomalies \cite{Minwalla:1997ka,Cordova:2016emh,Eberhardt:2020cxo}.  Celebrated examples include the two-dimensional $\mathcal{N}=1$ SCFT in the minimal model \cite{Friedan:1984rv} and the four-dimensional $\mathcal{N}=4$ super-Yang--Mills theory \cite{Brink:1976bc,Gliozzi:1976qd}, which is exactly conformal at any gauge coupling.

Although proposed over four decades ago, SUSY remains experimentally elusive in its original context of particle physics. 
Nevertheless, more recently, it has been proposed that SUSY may emerge at certain 3D quantum critical points in condensed matter systems, including surfaces of topological superconductors and insulators \cite{Grover:2013rc,Ponte:2012ru,Jian:2014pca,Witczak-Krempa:2015jca,Li:2016drh} and quantum phase transitions in Dirac materials \cite{Lee_2007,Li:2017dkj,Jian2017}. 
Unlike in two dimensions, however, exactly solvable or controllable microscopic realizations of interacting SCFTs in 3D remain scarce, and direct non-perturbative access to the emergent SCFT and its superconformal algebra continues to pose a major challenge.

\begin{figure}[t]
    \centering
    \begin{minipage}{0.23\textwidth}
        \centering
        \includegraphics[width=0.9\linewidth]{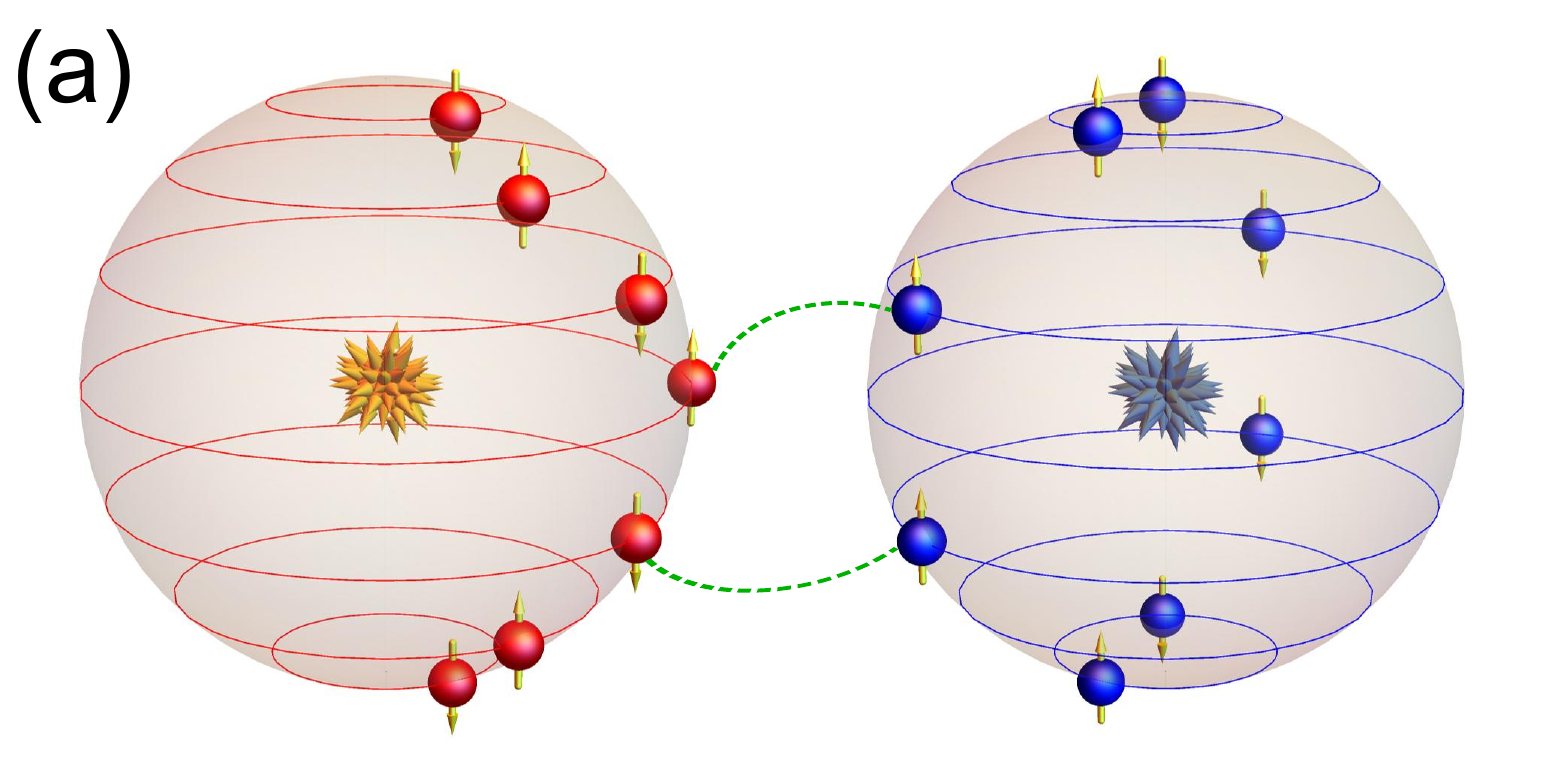}
        \\[1ex]
        \vspace{0.3ex} 
        \includegraphics[width=0.9\linewidth]{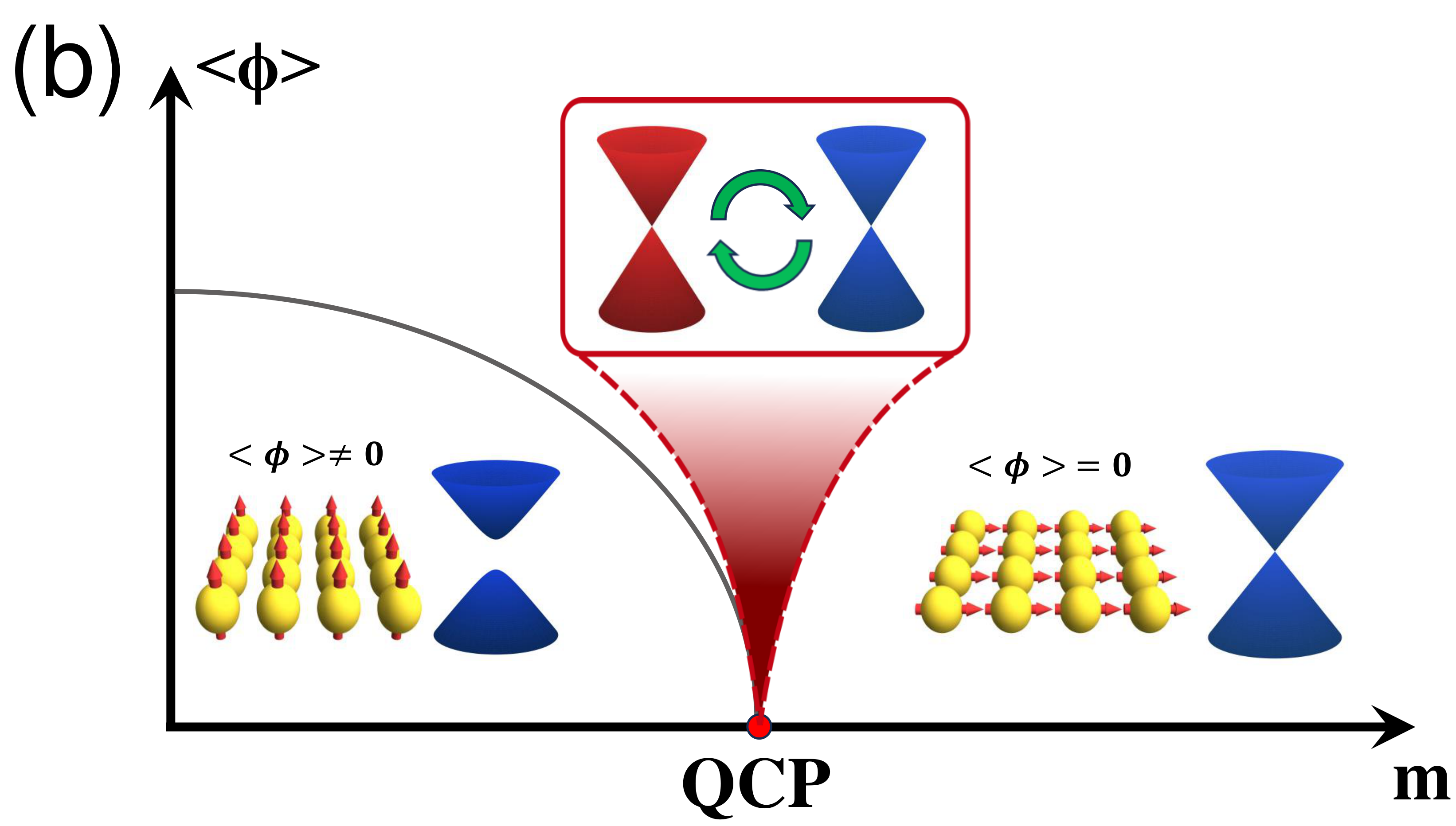}
    \end{minipage}
    \hfill
    \begin{minipage}{0.24\textwidth}
        \centering
        \includegraphics[width=1.0\linewidth]{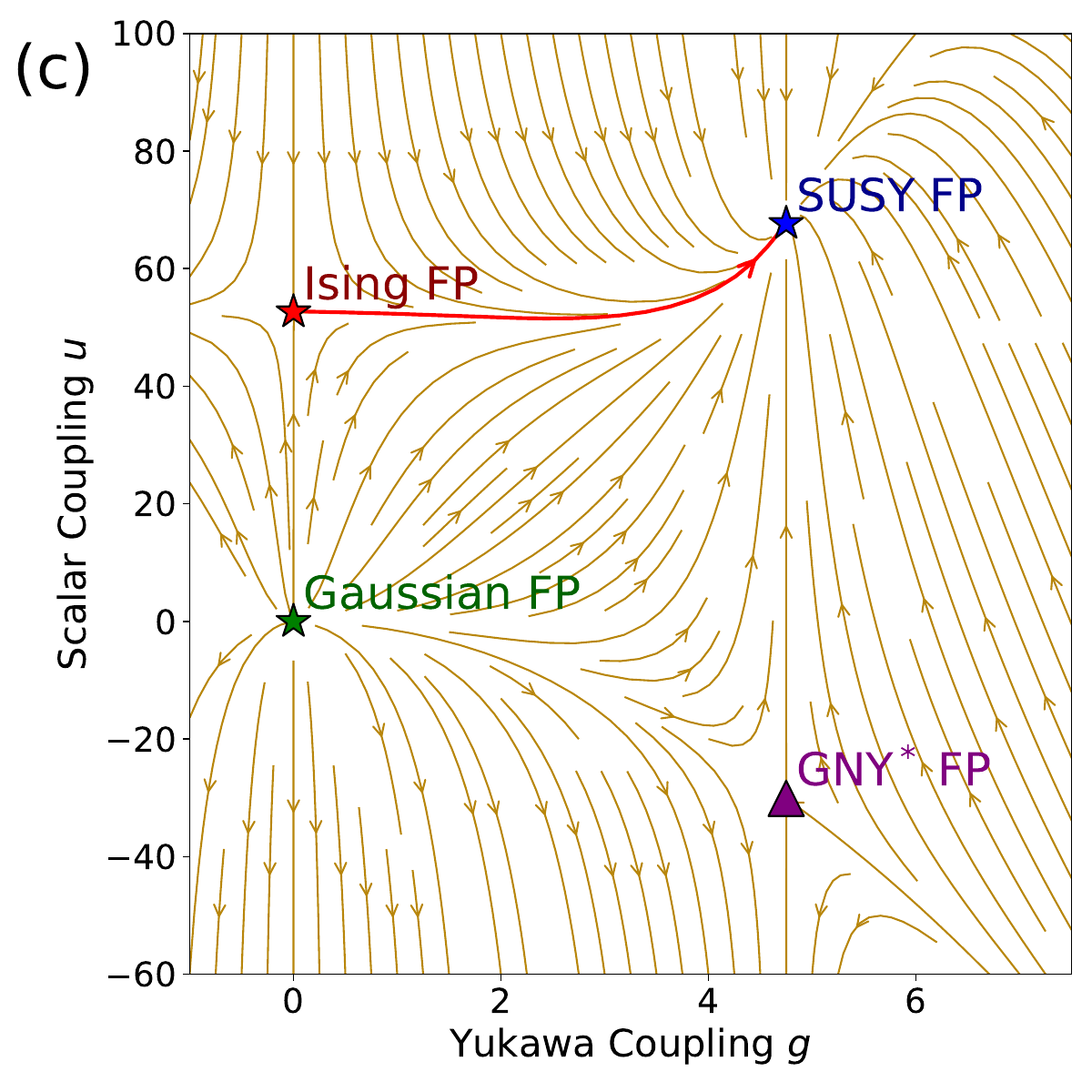}
    \end{minipage}
    \caption{(a) Our GNY model, Eq.~(\ref{eq:GNY}), consists of fermions (red dots) and bosons (blue dots), each residing on their own fuzzy sphere~\cite{Zhu:2022gjc}, and interacting with each other via the Yukawa interaction (green dashed line). (b)  A schematic phase diagram of the GNY model by tuning the bosonic field mass $m$, with a critical point separating a paramagnet with a massless fermion from a symmetry-breaking ferromagnet with a massive fermion. SUSY is expected to emerge at the quantum critical point, which exchanges bosonic and fermionic degrees of freedom. (c) RG phase diagram for the 3D $N=1$ GNY model (reproduced using $\beta$ function in \cite{Fei:2016sgs}), with the horizontal and vertical axes representing the Yukawa coupling $g$ between the boson and fermion and the   bosonic quartic coupling $u$, respectively. The diagram encloses a free Gaussian fixed point (green star), an Ising CFT fixed point (red star), an attractive SUSY CFT fixed point (blue star), and an unstable non-SUSY fixed point, also called GNY$^*$~\cite{Iliesiu:2015qra,Iliesiu:2017nrv}) (purple triangle). 
    }\label{fig:combined}
\end{figure}

In this paper, we investigate emergent 3D SCFTs through a microscopic realization of the Gross--Neveu--Yukawa (GNY) model with a single Majorana fermion flavor $N=1$ \cite{Gross:1974jv,Zinn-Justin:1991ksq,Fei:2016sgs} (we denote by $N$ the flavor of two-component Majorana fermions, while $\mathcal{N}$ is the SUSY index). Since the 3D $N=1$ GNY model is not accessible using conventional lattice regularizations, we gain non-perturbative access to the corresponding operator spectra by the fuzzy-sphere regularization~\cite{Zhu:2022gjc}, Fig.~\ref{fig:combined}(a), which preserves full rotational symmetry and allows direct application of the state--operator correspondence. At the quantum critical point, Fig.~\ref{fig:combined}(b), we observe signatures of  $\mathcal{N}=1$ superconformal invariance, evidenced by characteristic SUSY relations between fermionic and bosonic operators within the same supermultiplet. 
Furthermore, explicit operator spectral flows reveal the correspondence between operators of the decoupled CFTs and those of the interacting superconformal fixed point, see Fig.~\ref{fig:combined}(c). These results provide unbiased, non-perturbative evidence for emergent 3D SUSY conformal invariance in a microscopic model and establish a controlled route to studying 3D SCFTs.

{\sl Model.---}We consider the infrared (IR) fixed point of the following GNY Lagrangian:
\begin{equation}
	\mathcal{L}_{\rm GNY} = \frac{1}{2}(\partial\phi)^2 + i\psi\cancel{\partial}\psi + g\phi\overline{\psi}\psi + \frac{m}{2}\phi^2 + \frac{u}{4}\phi^4,
	\label{eq:GNY}
\end{equation}
where a single massless Majorana fermion (MF) $\psi$ couples to a $\mathbb{Z}_2$-odd scalar field $\phi$ via the Yukawa interaction with coupling strength $g$. The physics of the model in Eq.~\eqref{eq:GNY} has been studied extensively through perturbative renormalization group (RG) calculations \cite{Grover:2013rc,Fei:2016sgs,Zerf:2017zqi,Ihrig:2018hho,Mihaila:2017ble}. Perturbing the Ising fixed point by the Yukawa interaction, the RG flow eventually terminates at an IR fixed point described by $\mathcal{N}=1$ SCFT, Fig.~\ref{fig:combined}(c). When the bosonic field undergoes symmetry breaking, the Yukawa interaction gives mass to the fermion and breaks time-reversal parity simultaneously.
The resulting SCFT fixed point is also supported by modern conformal bootstrap techniques \cite{Iliesiu:2015qra,Iliesiu:2017nrv,Rong:2018okz,Atanasov:2018kqw,Atanasov:2022bpi,Erramilli:2022kgp,Mitchell:2024hix}.

At the SCFT fixed point, SUSY yields distinctive, highly constrained relations between the boson and fermion fields. Unfortunately, these relations have been challenging to directly verify in a microscopic setup. For example, quantum Monte Carlo studies  are impeded by severe fermion sign problems~\cite{Li:2014aoa}, while the anomaly of the Lagrangian in Eq.~\eqref{eq:GNY} under time-reversal parity obstructs its direct implementation in 3D local lattice model, leaving such numerical investigations currently underdeveloped~\cite{Li_sciadv2018}.
Additionally, the numerical bootstrap results show discrepancies with earlier Monte Carlo estimates of critical exponents at $N = 2$ \cite{Tabatabaei:2021tqv}, highlighting unresolved tensions between different methodologies.
These obstacles underscore the need for a controlled, non-perturbative, and unbiased microscopic approach to emergent SUSY in 3D, which we demonstrate below.

{\sl Method.---}Recently, the fuzzy sphere regularization~\cite{Zhu:2022gjc} has emerged as a powerful framework for studying 3D CFTs. This scheme has enabled non-perturbative investigations of diverse 3D CFTs, including $O(N)$ Wilson-Fisher criticality \cite{Han:2023lky,dey2025o3wilsonfisher,Lauchli_2025},  free theories \cite{He:2025ong,Taylor:2025odf,Voinea:2025iun,Zhou:2025kng}, deconfined criticality \cite{Zhou:2023qfi,Zhou:2024zud,yang2025o4},  Chern-Simons gauge theories \cite{Zhou:2025rmv}, and Lee-Yang criticality \cite{ArguelloCruz:2025zuq,Fan:2025bhc,EliasMiro:2025msj}. It also provides a platform for studying intrinsic CFT properties \cite{Hu:2023xak,Han:2023yyb,Hu:2024pen,Fan:2024vcz,Fardelli:2024qla} and their extensions such as defects \cite{Hu:2023ghk,Zhou:2023fqu,Cuomo:2024psk} and boundaries \cite{Zhou:2024dbt,Dedushenko:2024nwi}. Moreover, it inspires the study of interacting Dirac fermions on the spherical geometry \cite{Gao:2025vho}.  

Here we realize the GNY model, Eq.~\eqref{eq:GNY}, on the fuzzy sphere by coupling the 3D Ising CFT \cite{Zhu:2022gjc} with the 3D gauged Majorana fermion CFT \cite{Voinea:2025iun}. Denoting the corresponding Hamiltonians by $H_{\text{Ising}}$ and $H_{\text{Majorana}} $,  our GNY model is defined by the following Hamiltonian acting on their tensor product space:  
\begin{align}	\label{eq:ham}
	\notag H &= H_{\text{Ising}} \otimes \mathbb{I}_{\text{Majorana}} + \mathbb{I}_{\text{Ising}} \otimes  r H_{\text{Majorana}} + \int \mathbf{d}\boldsymbol{\Omega}_f  \mathbf{d}\boldsymbol{\Omega}_b \\
	& \times \left\{ n_f^z(\boldsymbol{\Omega}_f) \left[ \lambda^{z0}(\boldsymbol{\Omega}_{fb}) n_b^0(\boldsymbol{\Omega}_b) + \lambda^{zx}(\boldsymbol{\Omega}_{fb}) n_b^x(\boldsymbol{\Omega}_b) \right] \right\}.
\end{align}
The first two terms represent the independent Hamiltonians of the Ising and Majorana CFTs, with the parameter $r$ fixing their speed of light to be the same. The third term is a microscopic version of the Yukawa $g$-coupling in Eq.~(\ref{eq:GNY}). It is expressed in terms of
the Ising order parameter \(n_f^z \) and the MF field density operator \(n_b^{i=0,x,y,z} \), which depend on the spherical angles $\boldsymbol{\Omega}_f$ and $\boldsymbol{\Omega}_b$, respectively -- see the Supplementary Material (SM) for details~\cite{sm}. The two are coupled via a short-range interaction $\lambda^{ij}_M(\boldsymbol{\Omega}_f-\boldsymbol{\Omega}_b)\equiv \lambda^{ij}_M (\boldsymbol{\Omega}_{fb})$, which is expanded in terms of Haldane pseudopotentials parametrized by relative angular momentum $M$~\cite{Haldane1983}. We restrict to the following types of interaction terms: $\{\lambda\}=\{ \lambda_0^{z0},\lambda_1^{z0},\lambda_0^{zx},\lambda_1^{zx} \}$, involving only $M=0$ and $M=1$ pseudopotentials.
Setting all the couplings \(\{\lambda \}= 0\), we recover the product of 3D Ising CFT and a gauged MF CFT, Ising$\otimes$MF, shown by the red star in Fig.~\ref{fig:combined}(c). We note that  ``gauged'' means that the MF is coupled to a $\mathbb{Z}_2$ gauge field, which enforces a correspondence between the angular momentum sectors and the boson number (or magnetic flux) configurations on the fuzzy sphere~\cite{Voinea:2025iun}. Tuning \(\{\lambda\}\) is then expected to drive the system flow to a fixed point captured by the 3D \(\mathcal{N} = 1\) SCFT [blue star in Fig.~\ref{fig:combined} (c)]. 

The fuzzy sphere regularization introduces a monopole charge $s$ at the center of the sphere and confines the charged particles to the lowest Landau level (LLL), see Fig.~\ref{fig:combined}(a) and Refs.~\cite{Haldane1983,Zhu:2022gjc}. This setup naturally preserves the full \(\mathrm{SO(3)}\) spatial rotational symmetry -- a crucial advantage over lattice regularizations that typically break this symmetry down to discrete subgroups. Consequently, all eigenstates possess well-defined angular momentum quantum numbers $L$, corresponding directly to the Lorentz spins of the emergent CFT. 
The bosonic and fermionic fields, respectively, take integer and half-integer angular momenta.
The computational resources limit us to the numerically accessible system sizes up to  \(s=3\) (filled by \(N_b=8\) bosons, \(N_f=7\) fermions) for integer spin sectors, and \(s=5/2\) (\(N_b=7\), \(N_f=6\)) for half-integer spin sectors~\cite{sm}. 

\begin{figure}[t]
	\centering
	\includegraphics[width=0.48\textwidth]{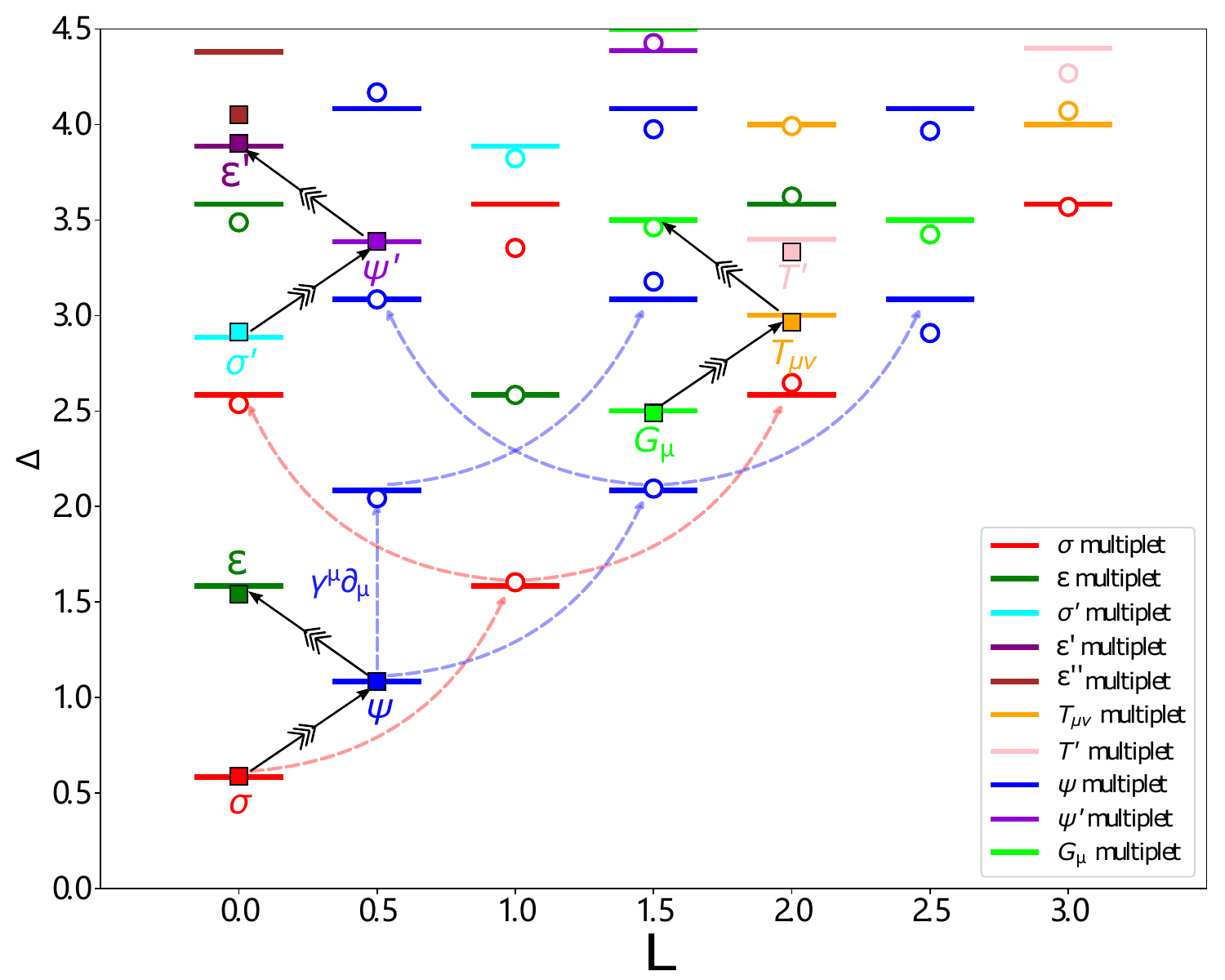}
	\caption{ The low-lying operator spectra for the 3D Ising SCFT. The squares and hollow circles respectively denote the conformal primary and descendant fields, obtained from our model Hamiltonian, Eq.~\eqref{eq:ham}, with $s = 3$ (containing integer angular momentum states) and $s = 5/2$ (containing half-integer angular momentum states). 
	The critical point $\{\lambda_c\}$ is determined by minimizing the associated cost function, which also sets the energy spectrum normalization \cite{sm}. 
	The expected CFT operators are indicated, along with their scaling dimensions computed via conformal bootstrap and marked by short horizontal lines \cite{Rong:2018okz,Atanasov:2018kqw,Atanasov:2022bpi,Erramilli:2022kgp}. 
	Different conformal multiplets are distinguished by color, with dashed lines indicating the actions of the conformal generators $P_\mu$ and $K_\mu$. Within the three observed supermultiplets, superconformal descendant states are connected by solid black arrows, which align with the emergent $\mathcal{N}=1$ superconformal algebra.  
	}
	\label{fig:spectrum}
\end{figure}

{\sl Operator spectra.---}A defining feature of the fuzzy sphere approach is that the conformal mapping \(\mathbb{S}^2 \times \mathbb{R} \to \mathbb{R}^3\) establishes a direct state-operator correspondence inherent to radial quantization on \(\mathbb{S}^2 \times \mathbb{R}\) \cite{Cardy:1984epx,Cardy:1985lth}. 
We firstly determine the critical point by minimizing cost functions incorporating low-lying primaries and their descendants (for details, see SM~\cite{sm}), which quantify deviations from the expected energy gap ratios of the \(\mathcal{N}=1\) supersymmetric Ising CFT. Parameter space scanning reveals a pronounced minimum of the cost function at \(\lambda_c^{s=3} = \{\lambda^{z0}_0 = 0.09, \lambda^{z0}_1 = 0.09, \lambda^{zx}_0 = 0.43, \lambda^{zx}_1 = -0.17\}\) for $s=3$, and \(\lambda_c^{s=5/2} = \{\lambda^{z0}_0 = 0.15, \lambda^{z0}_1 = 0.06, \lambda^{zx}_0 = 0.45, \lambda^{zx}_1 = -0.17\}\) for $s=5/2$. 
The resulting operator spectra shown in Fig. \ref{fig:spectrum} confirm the emergence of both conformal symmetry and SUSY at the critical point, as discussed below.

To begin with, the operator spectrum can be organized into conformal multiplets, indicating the emergence of conformal symmetry. Specifically, each primary field \( \mathcal{O}_{\text{primary}}\) and its descendants, \(\partial^\ell \Box^n \mathcal{O}_{\text{primary}}\), exhibit integer spacings in scaling dimensions. As shown in Fig.~\ref{fig:spectrum}, for the lowest scalar primary \(\sigma\) with scaling dimension \(\Delta_\sigma \approx 0.591\), its first descendant, namely \(\partial_\mu \sigma\), appears at \(\Delta_{\partial_\mu \sigma} \approx \Delta_\sigma+1\) ($L=1$), and the second-order descendants \(\partial_{\mu} \partial_{\nu} \sigma\), \(\Box \sigma\) at \(\Delta_{\Box \sigma} \approx \Delta_\sigma+2\) ($L=0$) and \(\Delta_{\partial_{\mu} \partial_{\nu} \sigma} \approx \Delta_\sigma+2\) ($L=2$), respectively, both aligning with the nearly perfect integer-spacing pattern predicted by conformal symmetry. The same integer-spacing structure applies to other conformal multiplets.

Within the range $\Delta\leq4.5$ and $L\leq3$ in our numerical spectrum, we have identified ten conformal multiplets, whose scaling dimensions and spin quantum numbers of the primary fields and their corresponding descendant fields are all consistent with the representations of 3D conformal algebra. In the $L=0$ sector, we found three relevant scalar primaries: $\sigma \sim \phi$, $\epsilon \sim \phi^2$, and $\sigma' \sim \phi^3$. $\sigma$ and $\sigma'$ are parity-odd. Therefore, while preserving time-reversal parity, the critical point can be reached in the IR by tuning only a single parameter in the microscopic model \cite{Grover:2013rc}. Among the fermionic fields, we identified three conformal multiplets. The first is the fundamental field $\psi$. In addition to the regular conformal descendants generated by acting with the conformal generators $P_{\mu}$ and $K^{\mu}$, this family includes a set of special descendants $\phi \psi$, due to the fermionic equation of motion in the GNY model:
\begin{equation}
\label{eq:fer_EOM}
	\gamma^\mu \partial_\mu \psi + g \phi \psi = 0.
\end{equation}
Thus, applying $\gamma^{\mu} \partial_{\mu}$ yields a set of model-specific descendant fields. Their Lorentz spin quantum numbers remain unchanged, while their scaling dimensions increase by 1 (indicated by the upward blue dashed lines in Fig.~\ref{fig:spectrum}). 
Crucially, in the $L=3/2$ sector, we identified a conserved current near the unitarity bound, $ G_{\mu}$, corresponding to the $\mathcal{N}=1$ supercurrent \cite{Erramilli:2022kgp}. In the $L=2$ sector, we found two spin-2 tensor operators. The lowest one corresponds to the energy-momentum tensor $T_{\mu\nu}$, while the other should correspond to the parity-even spin-2 operator  $T'_{\mu\nu}$  in another SUSY multiplet \cite{Atanasov:2018kqw}.
Tables~II and III of the SM~\cite{sm} list the obtained values of the scaling dimensions, while a quantitative comparison with the numerical bootstrap \cite{Iliesiu:2015qra,Iliesiu:2017nrv,Rong:2018okz,Atanasov:2018kqw,Atanasov:2022bpi,Erramilli:2022kgp,Mitchell:2024hix} is also presented. The discrepancies are consistently below \(3\%\) for low-lying primary fields within numerical uncertainties.

Remarkably, the operator spectrum further demonstrates that the bosonic and fermionic fields pair into supermultiplets. We have identified three supermultiplets by inspecting their related scaling dimensions and quantum numbers, marked as black solid arrows in Fig.~\ref{fig:spectrum}: (i) The leading bosonic scalar primary $\sigma$ \( (\Delta_\sigma \approx 0.591)\) and its superpartners, including the fermionic primary \(\psi\) (\(\Delta_\psi \approx 1.109\)) and the subleading bosonic scalar primary $\epsilon$ (\(\Delta_\epsilon \approx 1.516\)),  approximately satisfy the SUSY relation:
\begin{equation}
	\label{eq:susy}
	\Delta_\epsilon = \Delta_\psi + \frac{1}{2} = \Delta_\sigma + 1.
\end{equation} 
This encodes the fundamental principle of SUSY: the supercharge operator $Q$ with scaling dimension of $\pm \frac{1}{2}$ turns bosons into fermions and vice versa~\cite{Cordova:2016emh,Eberhardt:2020cxo}.
(ii) A similar relationship is also revealed for another supermultiplet $\sigma',\epsilon',\psi'$, which reads: \(\Delta_{\epsilon'}=\Delta_{\psi'}+\frac{1}{2} = \Delta_{\sigma'} + 1\). This set of primaries lies at relatively high energies, which are difficult to access via other methods.
(iii) A primary field with spin-$\frac{3}{2}$ indeed appears around $\Delta\sim 2.5$, which is identified as the supercurrent \(G_\mu\) of the SCFT.
The energy-momentum tensor \(T_{\mu\nu}\) is paired with  this spin-$\frac{3}{2}$ supercurrent \(G_\mu\), as superpartner. 
In summary, the detection of the supercurrent \(G_\mu\), along with the superconformal multiplet structure, constitutes robust numerical evidence for the emergent \(\mathcal{N}=1\) supersymmetric conformal invariance at the critical point.

\begin{figure}[t]
    \centering
    \begin{minipage}{0.4\textwidth}
        \centering
        \includegraphics[width=1.0\linewidth]{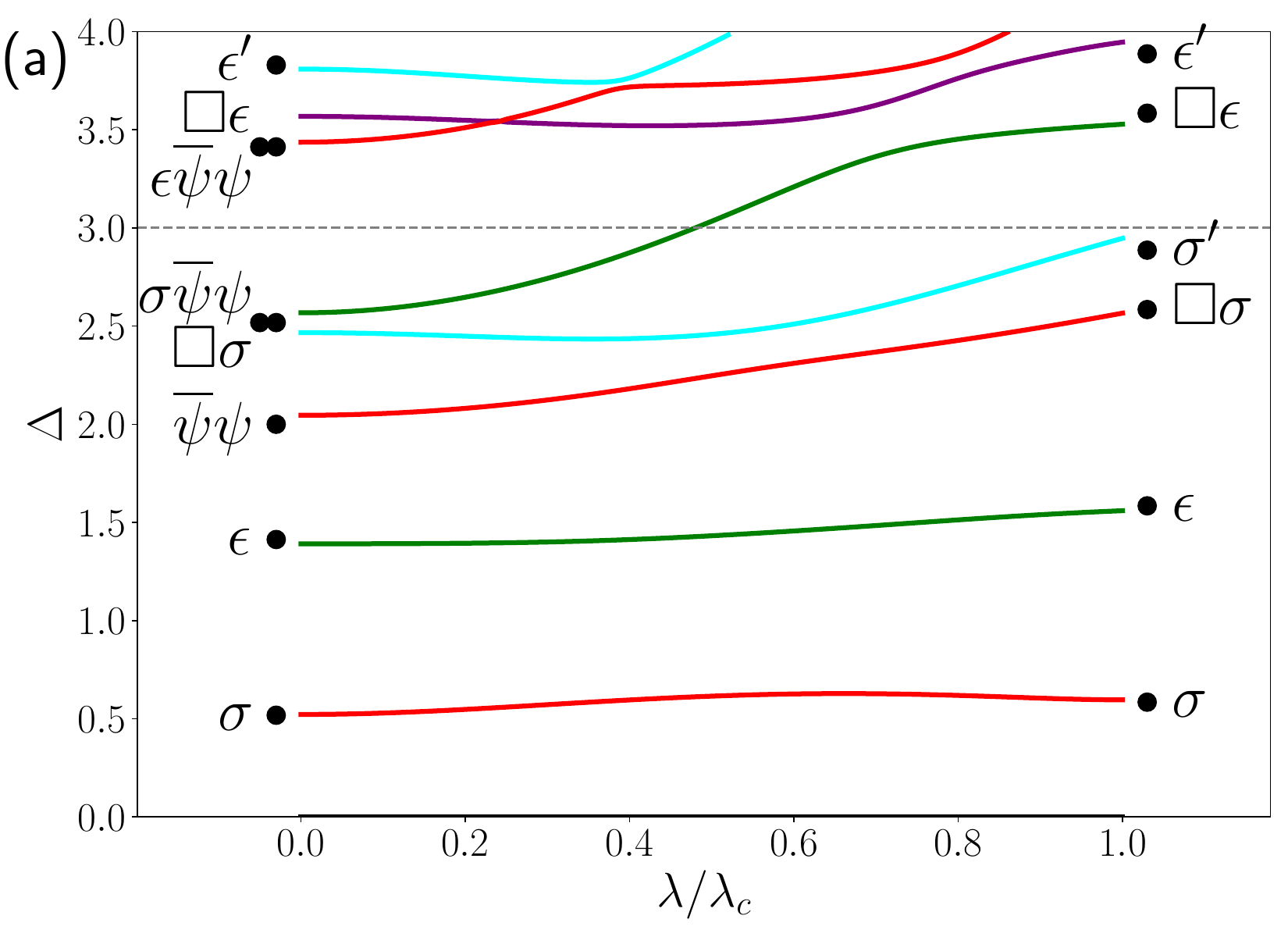}
    \end{minipage}
    \hfill
    \begin{minipage}{0.4\textwidth}
        \centering
        \includegraphics[width=1.0\linewidth]{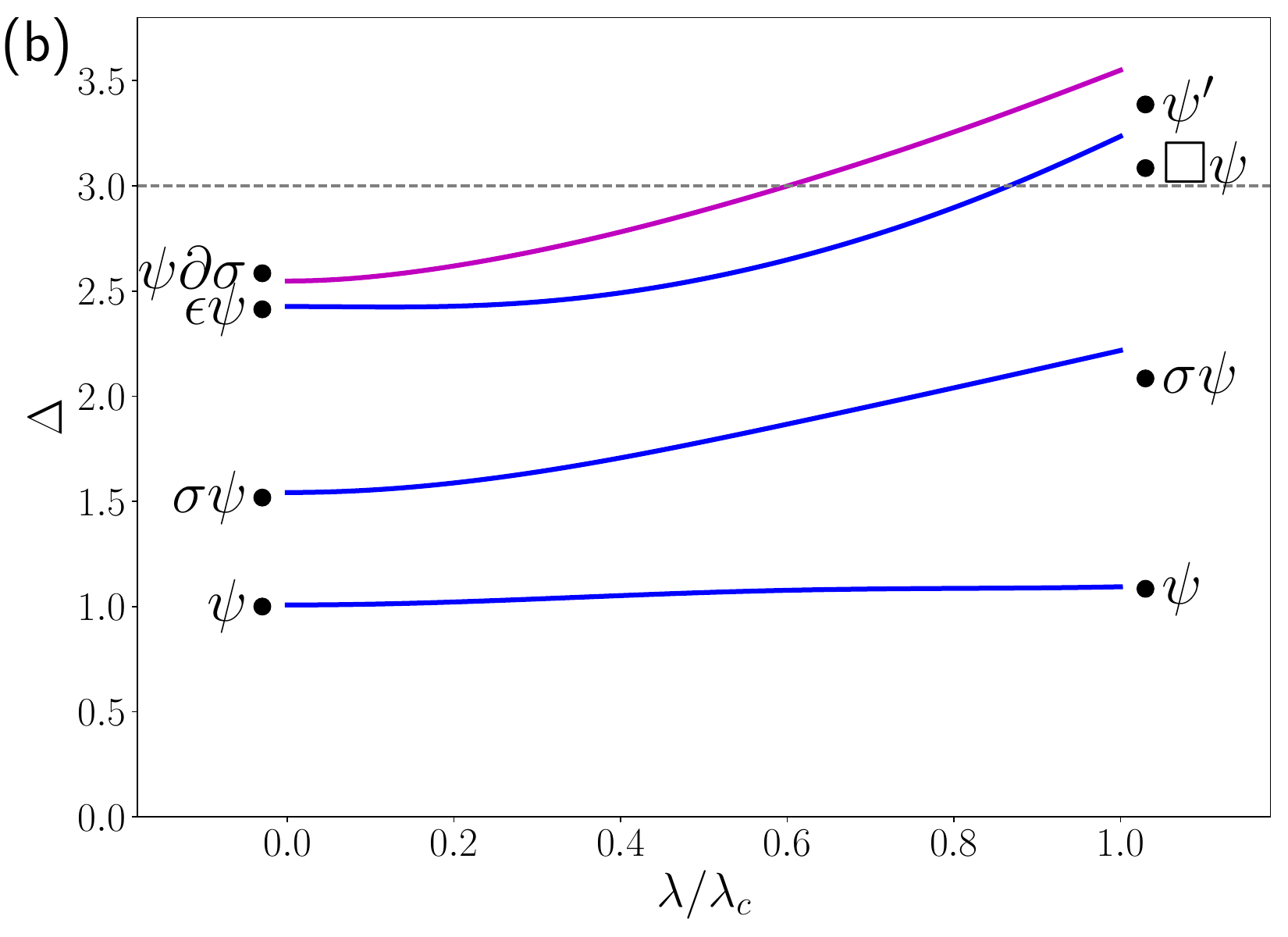}
    \end{minipage}
	\caption{Evolution of the low-lying spectra with varying  coupling $\{\lambda\}$. $\lambda/\lambda_c=0$ refers the 3D Ising$\otimes$MF CFT fixed point (red star in Fig.~\ref{fig:combined}) and $\lambda/\lambda_c=1$ represents the 3D Ising SCFT fixed point (blue star in Fig.~\ref{fig:combined}). The calculation is performed on the system size $s=3$ (a) and $s=5/2$ (b). We renormalize the spectra  by setting $\Delta_T=3$ for any $\{\lambda\}$. For half-integer momentum sectors, the ground state $E_{GS}$ and $E_T$ are approximated by the average values of the two adjacent integer momentum sectors \cite{Voinea:2025iun} (see text).  
    } 
	\label{fig:RG}
\end{figure}

{\sl Operator flow.---}The operator content of the SCFT is further manifested in the RG flows. 
Figure~\ref{fig:RG} depicts the evolution of the operator spectrum  from the decoupled Ising$\otimes$MF fixed point to that of the emergent SCFT fixed point, by tuning the Yukawa coupling strength $\{\lambda\}$.   
We emphasize that while any point along this trajectory flows to the GNY fixed point in the IR limit, finite-size effects and UV cutoffs cause different values of $\{\lambda\}$ to represent distinct RG stages.

For bosonic fields, the Ising primaries $\sigma$ and $\epsilon$ evolve continuously into their GNY counterparts. Concurrently, the MF mass operator $\bar{\psi}\psi$ exhibits mixing with the descendant field $\Box\sigma=\Box \phi$ along the flow trajectory. These operators ultimately flow to the GNY operators $\sigma' \sim \phi^3$ and $\Box\sigma$ according to the equation of motion from Eq.~\eqref{eq:GNY}:
\begin{equation}
	\Box\phi + u \phi^3 + g \bar{\psi}\psi = 0,
\end{equation}
which enforces constraints among operators at criticality and also matches the $\epsilon-$expansion prediction~\cite{Fei:2016sgs}. Meanwhile, the Yukawa coupling operator $\sigma\bar{\psi}\psi$ evolves from relevant, near the Ising$\otimes$MF fixed point, to irrelevant at the GNY fixed point, directly resulting from the UV to IR RG flows. This provides further evidence for the emergent SUSY of the GNY model.
The original descendant field $\Box \epsilon$ flows to $\epsilon'$ and the composite operator $\epsilon \bar{\psi}\psi$ flows to $\Box^2 \sigma$, both of which preserve the parity charge.

Fermionic operators display analogous RG patterns. The fundamental fermion field $\psi$ flows from its free-field dimension $\Delta_\psi=1$ to $\Delta_\psi \approx 1.109$ at the SCFT fixed point, consistent with the superconformal relation in Eq.~\eqref{eq:susy}. More notably, the composite operator $\sigma\psi$ transforms from a combined primary operator $\Delta_{\sigma\psi} \approx \Delta_\sigma + \Delta_\psi$ at the Ising$\otimes$MF fixed point into a descendant of $\psi$ at GNY fixed point, acquiring the dimension $\Delta_{\sigma\psi} = \Delta_\psi + 1$, governed by the fermionic equation of motion Eq.~\eqref{eq:fer_EOM}, which implies $\sigma\psi \propto \gamma^{\mu}\partial_\mu\psi$ at the IR fixed point. In light of the above, the RG running process reveals the operator correspondence between the decoupled Ising$\otimes$MF CFT and SCFT.

{\sl Summary and discussion.---}We have demonstrated a non-perturbative framework for probing the emergence of 3D SCFTs in unbiased microscopic models. We illustrated our approach using the $\mathcal{N}=1$ supersymmetric Ising CFT, directly verifying the emergent superconformal symmetry through identification of the superconformal multiplets. Crucially, the scaling dimensions of low-lying fields satisfy the SUSY relations, such as Eq.~(\ref{eq:susy}), with high precision, while a protected supercurrent around $\Delta_{G_\mu} = 2.5$ is also observed. These results confirm the emergence of $\mathcal{N}=1$ super-Ising universality class by enabling non-perturbative determination of operator dimensions explicitly, establishing our approach as a powerful tool for strongly-coupled fixed points hard to access with conventional lattice approaches~\cite{Lee:2006if,Grover:2013rc}. Moreover, the same methodology also enables direct tracking of the evolution for low-lying operators across criticality, revealing how SUSY multiplets emerge from non-supersymmetric UV frameworks.

Our results open several new directions for future studies of 3D SCFTs. For example, our model enables computation of additional universal data in the 3D super-Ising CFT \cite{Fei:2016sgs,Iliesiu:2017nrv,Diab:2016spb,Giombi:2017rhm,Goykhman:2020tsk,Rong:2018okz,Atanasov:2018kqw,Atanasov:2022bpi,Erramilli:2022kgp}, including operator product expansion coefficients and central charges—particularly $C_T$, which would provide stringent tests of superconformal algebra and bootstrap results. This also allows the computation of the universal finite $F-$term in the entanglement entropy \cite{Hu:2024pen}, for which the RG flow between decoupled fixed points and GNY fixed points provides an ideal platform \cite{Fei:2016sgs,Jafferis:2011zi,Klebanov:2011gs,Closset:2012vg,Casini:2012ei,Giombi:2014xxa}. Another natural extension of our construction is to multiple species of bosons and fermions, which may enable the realization of a wider class of boson-fermion coupled theories \cite{Nambu:1961tp,Liendo:2021wpo,Pannell:2023tzc,Jack:2024sjr}, including the $\mathcal{N}=2$ SCFTs, where the enhanced supersymmetry imposes richer constraints on the spectrum \cite{Strassler:2003qg,Lee:2006if,Bobev:2015vsa,Bobev:2015jxa,Li:2017kck}.
 Moreover, our framework may also allow one to examine the existence and unitarity of the 3D GNY$^*$ CFT \cite{Fei:2016sgs,Iliesiu:2017nrv,Nakayama:2022svf} by perturbing the Ising fuzzy sphere \cite{Taylor:2025odf}.
Extended defects provide yet another arena of novel aspects of fermionic CFTs and supersymmetric boundary conditions \cite{Giombi:2021cnr,Giombi:2022vnz,Herzog:2022jlx,Barrat:2023ivo,Diatlyk:2024zkk,Jiang:2025sfb,Giombi:2025evu}. Finally, incorporating additional perturbations would facilitate non-perturbative studies of 3D supersymmetric and non-supersymmetric field theories. These advances, combined with refinements in bootstrap techniques and diagrammatics methods, may lead to a more comprehensive understanding of the landscape of 3D quantum field theories.

\begin{acknowledgments}
\textit{Acknowledgments.---}We thank Yin-Chen He, Ning Su, Hong Yao, Yue Yu, Xinyu Zhang, Zheng Zhou for useful discussion. Y.T., L.D.H. and W.Z. were supported by NSFC under No. 12474144. 
C.V. and Z.P. acknowledge support by the Leverhulme Trust Research Leadership Award RL-2019-015 and EPSRC Grants EP/Z533634/1, UKRI1337. Statement of compliance with EPSRC policy framework on research data: This publication is theoretical work that does not require supporting research data. This research was supported in part by grant NSF PHY-2309135 to the Kavli Institute for Theoretical Physics (KITP). Computational portions of this research have made use of DiagHam~\cite{diagham} and FuzzifiED~\cite{FuzzifiED} software libraries, and they were carried out on Aire, part of the High-Performance Computing facilities at the University of Leeds.
\end{acknowledgments}

\bibliography{ref}

@article{sm,
	Journal = {Supplementary material}}

@article{Sohnius:1985qm,
    author = "Sohnius, M. F.",
    title = "{Introducing Supersymmetry}",
    doi = "10.1016/0370-1573(85)90023-7",
    journal = "Phys. Rept.",
    volume = "128",
    pages = "39--204",
    year = "1985"
}

@book{Wess_Bagger_Book,
    author = "Wess, J. and Bagger, J.",
    title = "{Supersymmetry and supergravity}",
    isbn = "978-0-691-02530-8",
    publisher = "Princeton University Press",
    address = "Princeton, NJ, USA",
    year = "1992"
}

@book{Weinberg:2000cr,
    author = "Weinberg, Steven",
    title = "{The quantum theory of fields. Vol. 3: Supersymmetry}",
    isbn = "978-0-521-67055-5, 978-1-139-63263-8, 978-0-521-67055-5",
    publisher = "Cambridge University Press",
    month = "6",
    year = "2013"
}

@book{Cardy_book,
	title = {Scaling and Renormalization in Statistical Physics},
	author = {Cardy, J.},
	isbn = {9781316036440},
	year = {1996},
	publisher = {Cambridge University Press, Cambridge, England}
}

@book{yellowbook,
    title = {Conformal Field Theory},
    author = {Philippe Francesco, Pierre Mathieu, David Sénéchal},
    isbn = {978-1-4612-2256-9},
    series = {Graduate Texts in Contemporary Physics},
    year = {1997},
    publisher = {Springer New York, NY}
}

@article{Minwalla:1997ka,
    author = "Minwalla, Shiraz",
    title = "{Restrictions imposed by superconformal invariance on quantum field theories}",
    eprint = "hep-th/9712074",
    archivePrefix = "arXiv",
    reportNumber = "PUPT-1748",
    doi = "10.4310/ATMP.1998.v2.n4.a4",
    journal = "Adv. Theor. Math. Phys.",
    volume = "2",
    pages = "783--851",
    year = "1998"
}

@article{Cordova:2016emh,
    author = "Cordova, Clay and Dumitrescu, Thomas T. and Intriligator, Kenneth",
    title = "{Multiplets of Superconformal Symmetry in Diverse Dimensions}",
    eprint = "1612.00809",
    archivePrefix = "arXiv",
    primaryClass = "hep-th",
    doi = "10.1007/JHEP03(2019)163",
    journal = "JHEP",
    volume = "03",
    pages = "163",
    year = "2019"
}

@article{Eberhardt:2020cxo,
    author = "Eberhardt, Lorenz",
    title = "{Superconformal symmetry and representations}",
    eprint = "2006.13280",
    archivePrefix = "arXiv",
    primaryClass = "hep-th",
    doi = "10.1088/1751-8121/abd7b3",
    journal = "J. Phys. A",
    volume = "54",
    number = "6",
    pages = "063002",
    year = "2021"
}

@inproceedings{Strassler:2003qg,
    author = "Strassler, Matthew J.",
    title = "{An Unorthodox introduction to supersymmetric gauge theory}",
    booktitle = "{Theoretical Advanced Study Institute in Elementary Particle Physics (TASI 2001): Strings, Branes and EXTRA Dimensions}",
    eprint = "hep-th/0309149",
    archivePrefix = "arXiv",
    reportNumber = "UW-PT-03-18",
    doi = "10.1142/9789812702821_0011",
    pages = "561--638",
    month = "9",
    year = "2003"
}

@article{Brink:1976bc,
    author = "Brink, Lars and Schwarz, John H. and Scherk, Joel",
    title = "{Supersymmetric Yang-Mills Theories}",
    reportNumber = "CALT-68-574",
    doi = "10.1016/0550-3213(77)90328-5",
    journal = "Nucl. Phys. B",
    volume = "121",
    pages = "77--92",
    year = "1977"
}

@article{Gliozzi:1976qd,
    author = "Gliozzi, F. and Scherk, Joel and Olive, David I.",
    title = "{Supersymmetry, Supergravity Theories and the Dual Spinor Model}",
    reportNumber = "CERN-TH-2253",
    doi = "10.1016/0550-3213(77)90206-1",
    journal = "Nucl. Phys. B",
    volume = "122",
    pages = "253--290",
    year = "1977"
}

@article{Friedan:1984rv,
    author = "Friedan, Daniel and Qiu, Zong-an and Shenker, Stephen H.",
    title = "{Superconformal Invariance in Two-Dimensions and the Tricritical Ising Model}",
    reportNumber = "EFI-84-35-CHICAGO",
    doi = "10.1016/0370-2693(85)90819-6",
    journal = "Phys. Lett. B",
    volume = "151",
    pages = "37--43",
    year = "1985"
}

@article{Gross:1974jv,
    author = "Gross, David J. and Neveu, Andre",
    title = "{Dynamical Symmetry Breaking in Asymptotically Free Field Theories}",
    reportNumber = "COO-2220-19",
    doi = "10.1103/PhysRevD.10.3235",
    journal = "Phys. Rev. D",
    volume = "10",
    pages = "3235",
    year = "1974"
}

@article{Zinn-Justin:1991ksq,
    author = "Zinn-Justin, Jean",
    title = "{Four fermion interaction near four-dimensions}",
    reportNumber = "SACLAY-SPH-T-91-092",
    doi = "10.1016/0550-3213(91)90043-W",
    journal = "Nucl. Phys. B",
    volume = "367",
    pages = "105--122",
    year = "1991"
}

@article{Iliesiu:2015qra,
    author = "Iliesiu, Luca and Kos, Filip and Poland, David and Pufu, Silviu S. and Simmons-Duffin, David and Yacoby, Ran",
    title = "{Bootstrapping 3D Fermions}",
    eprint = "1508.00012",
    archivePrefix = "arXiv",
    primaryClass = "hep-th",
    reportNumber = "PUPT-2480",
    doi = "10.1007/JHEP03(2016)120",
    journal = "JHEP",
    volume = "03",
    pages = "120",
    year = "2016"
}

@article{Iliesiu:2017nrv,
    author = "Iliesiu, Luca and Kos, Filip and Poland, David and Pufu, Silviu S. and Simmons-Duffin, David",
    title = "{Bootstrapping 3D Fermions with Global Symmetries}",
    eprint = "1705.03484",
    archivePrefix = "arXiv",
    primaryClass = "hep-th",
    reportNumber = "PUPT-2524",
    doi = "10.1007/JHEP01(2018)036",
    journal = "JHEP",
    volume = "01",
    pages = "036",
    year = "2018"
}

@article{Rong:2018okz,
    author = "Rong, Junchen and Su, Ning",
    title = "{Bootstrapping the minimal $ \mathcal{N} $ = 1 superconformal field theory in three dimensions}",
    eprint = "1807.04434",
    archivePrefix = "arXiv",
    primaryClass = "hep-th",
    doi = "10.1007/JHEP06(2021)154",
    journal = "JHEP",
    volume = "06",
    pages = "154",
    year = "2021"
}

@article{Atanasov:2018kqw,
    author = "Atanasov, Alexander and Hillman, Aaron and Poland, David",
    title = "{Bootstrapping the Minimal 3D SCFT}",
    eprint = "1807.05702",
    archivePrefix = "arXiv",
    primaryClass = "hep-th",
    doi = "10.1007/JHEP11(2018)140",
    journal = "JHEP",
    volume = "11",
    pages = "140",
    year = "2018"
}

@article{Atanasov:2022bpi,
    author = "Atanasov, Alexander and Hillman, Aaron and Poland, David and Rong, Junchen and Su, Ning",
    title = "{Precision bootstrap for the $ \mathcal{N} $ = 1 super-Ising model}",
    eprint = "2201.02206",
    archivePrefix = "arXiv",
    primaryClass = "hep-th",
    doi = "10.1007/JHEP08(2022)136",
    journal = "JHEP",
    volume = "08",
    pages = "136",
    year = "2022"
}

@article{Erramilli:2022kgp,
    author = "Erramilli, Rajeev S. and Iliesiu, Luca V. and Kravchuk, Petr and Liu, Aike and Poland, David and Simmons-Duffin, David",
    title = "{The Gross-Neveu-Yukawa archipelago}",
    eprint = "2210.02492",
    archivePrefix = "arXiv",
    primaryClass = "hep-th",
    reportNumber = "CALT-TH 2022-027",
    doi = "10.1007/JHEP02(2023)036",
    journal = "JHEP",
    volume = "02",
    pages = "036",
    year = "2023"
}

@article{Mitchell:2024hix,
    author = "Mitchell, Matthew S. and Poland, David",
    title = "{Bounding irrelevant operators in the 3d Gross-Neveu-Yukawa CFTs}",
    eprint = "2406.12974",
    archivePrefix = "arXiv",
    primaryClass = "hep-th",
    doi = "10.1007/JHEP09(2024)134",
    journal = "JHEP",
    volume = "09",
    pages = "134",
    year = "2024"
}

@article{Fei:2016sgs,
    author = "Fei, Lin and Giombi, Simone and Klebanov, Igor R. and Tarnopolsky, Grigory",
    title = "{Yukawa CFTs and Emergent Supersymmetry}",
    eprint = "1607.05316",
    archivePrefix = "arXiv",
    primaryClass = "hep-th",
    reportNumber = "PUPT-2504",
    doi = "10.1093/ptep/ptw120",
    journal = "PTEP",
    volume = "2016",
    number = "12",
    pages = "12C105",
    year = "2016"
}

@article{Grover:2013rc,
    author = "Grover, Tarun and Sheng, D. N. and Vishwanath, Ashvin",
    title = "{Emergent Space-Time Supersymmetry at the Boundary of a Topological Phase}",
    eprint = "1301.7449",
    archivePrefix = "arXiv",
    primaryClass = "cond-mat.str-el",
    doi = "10.1126/science.1248253",
    journal = "Science",
    volume = "344",
    number = "6181",
    pages = "280--283",
    year = "2014"
}

@article{Lee_2007,
	title={Emergence of supersymmetry at a critical point of a lattice model},
	volume={76},
	ISSN={1550-235X},
	url={http://dx.doi.org/10.1103/PhysRevB.76.075103},
	DOI={10.1103/physrevb.76.075103},
	number={7},
	journal={Physical Review B},
	publisher={American Physical Society (APS)},
	author={Lee, Sung-Sik},
	year={2007},
	month=aug }

@article{Li:2016drh,
    author = "Li, Zi-Xiang and Jiang, Yi-Fan and Yao, Hong",
    title = "{Edge quantum criticality and emergent supersymmetry in topological phases}",
    eprint = "1610.04616",
    archivePrefix = "arXiv",
    primaryClass = "cond-mat.str-el",
    doi = "10.1103/PhysRevLett.119.107202",
    journal = "Phys. Rev. Lett.",
    volume = "119",
    number = "10",
    pages = "107202",
    year = "2017"
}

@article{Jian:2014pca,
    author = "Jian, Shao-Kai and Jiang, Yi-Fan and Yao, Hong",
    title = "{Emergent Spacetime Supersymmetry in 3D Weyl Semimetals and 2D Dirac Semimetals}",
    eprint = "1407.4497",
    archivePrefix = "arXiv",
    primaryClass = "cond-mat.str-el",
    doi = "10.1103/PhysRevLett.114.237001",
    journal = "Phys. Rev. Lett.",
    volume = "114",
    number = "23",
    pages = "237001",
    year = "2015"
}

@article{Jian2017,
	title = {Emergence of Supersymmetric Quantum Electrodynamics},
	author = {Jian, Shao-Kai and Lin, Chien-Hung and Maciejko, Joseph and Yao, Hong},
	journal = {Phys. Rev. Lett.},
	volume = {118},
	issue = {16},
	pages = {166802},
	numpages = {6},
	year = {2017},
	month = {Apr},
	publisher = {American Physical Society},
	doi = {10.1103/PhysRevLett.118.166802},
	url = {https://link.aps.org/doi/10.1103/PhysRevLett.118.166802}
}

@article{Witczak-Krempa:2015jca,
    author = "Witczak-Krempa, William and Maciejko, Joseph",
    title = "{Optical conductivity of topological surface states with emergent supersymmetry}",
    eprint = "1510.06397",
    archivePrefix = "arXiv",
    primaryClass = "cond-mat.str-el",
    doi = "10.1103/PhysRevLett.116.100402",
    journal = "Phys. Rev. Lett.",
    volume = "116",
    number = "10",
    pages = "100402",
    year = "2016",
    note = "[Addendum: Phys.Rev.Lett. 117, 149903 (2016)]"
}

@article{Li:2017dkj,
author = {Zi-Xiang Li  and Abolhassan Vaezi  and Christian B. Mendl  and Hong Yao },
title = {Numerical observation of emergent spacetime supersymmetry at quantum criticality},
journal = {Science Advances},
volume = {4},
number = {11},
pages = {eaau1463},
year = {2018},
doi = {10.1126/sciadv.aau1463},
URL = {https://www.science.org/doi/abs/10.1126/sciadv.aau1463},
eprint = {https://www.science.org/doi/pdf/10.1126/sciadv.aau1463}}

@article{Li_sciadv2018,
   title={Numerical observation of emergent spacetime supersymmetry at quantum criticality},
   volume={4},
   ISSN={2375-2548},
   url={http://dx.doi.org/10.1126/sciadv.aau1463},
   DOI={10.1126/sciadv.aau1463},
   number={11},
   journal={Science Advances},
   publisher={American Association for the Advancement of Science (AAAS)},
   author={Li, Zi-Xiang and Vaezi, Abolhassan and Mendl, Christian B. and Yao, Hong},
   year={2018},
   month=nov }

@article{Mihaila:2017ble,
    author = "Mihaila, Luminita N. and Zerf, Nikolai and Ihrig, Bernhard and Herbut, Igor F. and Scherer, Michael M.",
    title = "{Gross-Neveu-Yukawa model at three loops and Ising critical behavior of Dirac systems}",
    eprint = "1703.08801",
    archivePrefix = "arXiv",
    primaryClass = "cond-mat.str-el",
    doi = "10.1103/PhysRevB.96.165133",
    journal = "Phys. Rev. B",
    volume = "96",
    number = "16",
    pages = "165133",
    year = "2017"
}

@article{Li:2014aoa,
    author = "Li, Zi-Xiang and Jiang, Yi-Fan and Yao, Hong",
    title = "{Fermion-sign-free Majarana-quantum-Monte-Carlo studies of quantum critical phenomena of Dirac fermions in two dimensions}",
    eprint = "1411.7383",
    archivePrefix = "arXiv",
    primaryClass = "cond-mat.str-el",
    doi = "10.1088/1367-2630/17/8/085003",
    journal = "New J. Phys.",
    volume = "17",
    number = "8",
    pages = "085003",
    year = "2015"
}

@article{Tabatabaei:2021tqv,
    author = "Tabatabaei, S. Mojtaba and Negari, Amir-Reza and Maciejko, Joseph and Vaezi, Abolhassan",
    title = "{Chiral Ising Gross-Neveu Criticality of a Single Dirac Cone: A Quantum Monte~Carlo Study}",
    eprint = "2112.09209",
    archivePrefix = "arXiv",
    primaryClass = "cond-mat.str-el",
    doi = "10.1103/PhysRevLett.128.225701",
    journal = "Phys. Rev. Lett.",
    volume = "128",
    number = "22",
    pages = "225701",
    year = "2022"
}

@article{Zhu:2022gjc,
    author = "Zhu, Wei and Han, Chao and Huffman, Emilie and Hofmann, Johannes S. and He, Yin-Chen",
    title = "{Uncovering Conformal Symmetry in the 3D Ising Transition: State-Operator Correspondence from a Quantum Fuzzy Sphere Regularization}",
    eprint = "2210.13482",
    archivePrefix = "arXiv",
    primaryClass = "cond-mat.stat-mech",
    doi = "10.1103/PhysRevX.13.021009",
    journal = "Phys. Rev. X",
    volume = "13",
    number = "2",
    pages = "021009",
    year = "2023"
}

@article{Haldane1983,
  title = {Fractional Quantization of the Hall Effect: A Hierarchy of Incompressible Quantum Fluid States},
  author = {Haldane, F. D. M.},
  journal = {Phys. Rev. Lett.},
  volume = {51},
  issue = {7},
  pages = {605--608},
  numpages = {0},
  year = {1983},
  month = {Aug},
  publisher = {American Physical Society},
  doi = {10.1103/PhysRevLett.51.605},
  url = {https://link.aps.org/doi/10.1103/PhysRevLett.51.605}
}

@article{Cardy:1984epx,
    author = "Cardy, J. L.",
    title = "{Conformal invariance and universality in finite-size scaling}",
    doi = "10.1088/0305-4470/17/7/003",
    journal = "J. Phys. A",
    volume = "17",
    number = "7",
    pages = "L385",
    year = "1984"
}

@article{Cardy:1985lth,
    author = "Cardy, J. L.",
    title = "{Universal amplitudes in finite-size scaling: generalisation to arbitrary dimensionality}",
    doi = "10.1088/0305-4470/18/13/005",
    journal = "J. Phys. A",
    volume = "18",
    number = "13",
    pages = "L757--L760",
    year = "1985"
}

@article{Lauchli_2025,
   title={Exact diagonalization, matrix product states and conformal perturbation theory study of a 3D Ising fuzzy sphere model},
   volume={19},
   ISSN={2542-4653},
   url={http://dx.doi.org/10.21468/SciPostPhys.19.3.076},
   DOI={10.21468/scipostphys.19.3.076},
   number={3},
   journal={SciPost Physics},
   publisher={Stichting SciPost},
   author={Läuchli, Andreas and Herviou, Loïc and Wilhelm, Patrick and Rychkov, Slava},
   year={2025},
   month=sep }

@article{Zhou:2023qfi,
    author = "Zhou, Zheng and Hu, Liangdong and Zhu, W. and He, Yin-Chen",
    title = "{SO(5) Deconfined Phase Transition under the Fuzzy-Sphere Microscope: Approximate Conformal Symmetry, Pseudo-Criticality, and Operator Spectrum}",
    eprint = "2306.16435",
    archivePrefix = "arXiv",
    primaryClass = "cond-mat.str-el",
    doi = "10.1103/PhysRevX.14.021044",
    journal = "Phys. Rev. X",
    volume = "14",
    number = "2",
    pages = "021044",
    year = "2024"
}

@misc{yang2025o4,
      title={Conformal Operator Flows of the Deconfined Quantum Criticality from $\mathrm{SO}(5)$ to $\mathrm{O}(4)$}, 
      author={Shuai Yang and Liang-dong Hu and Chao Han and W. Zhu and Yan Chen},
      year={2025},
      eprint={2507.01322},
      archivePrefix={arXiv},
      primaryClass={cond-mat.str-el},
      url={https://arxiv.org/abs/2507.01322}, 
}

@article{Han:2023lky,
    author = "Han, Chao and Hu, Liangdong and Zhu, W.",
    title = "{Conformal operator content of the Wilson-Fisher transition on fuzzy sphere bilayers}",
    eprint = "2312.04047",
    archivePrefix = "arXiv",
    primaryClass = "cond-mat.str-el",
    doi = "10.1103/PhysRevB.110.115113",
    journal = "Phys. Rev. B",
    volume = "110",
    number = "11",
    pages = "115113",
    year = "2024"
}

@misc{dey2025o3wilsonfisher,
      title={Conformal Data for the O(3) Wilson-Fisher CFT from Fuzzy Sphere Realization of Quantum Rotor Model}, 
      author={Arjun Dey and Loic Herviou and Christopher Mudry and Andreas Martin Läuchli},
      year={2025},
      eprint={2510.09755},
      archivePrefix={arXiv},
      primaryClass={cond-mat.str-el},
      url={https://arxiv.org/abs/2510.09755}, 
}

@article{Zhou:2024zud,
    author = "Zhou, Zheng and He, Yin-Chen",
    title = "{3D Conformal Field Theories with Sp(N) Global Symmetry on a Fuzzy Sphere}",
    eprint = "2410.00087",
    archivePrefix = "arXiv",
    primaryClass = "hep-th",
    doi = "10.1103/xstj-xvcy",
    journal = "Phys. Rev. Lett.",
    volume = "135",
    number = "2",
    pages = "026504",
    year = "2025"
}

@article{Zhou:2025rmv,
    author = "Zhou, Zheng and Wang, Chong and He, Yin-Chen",
    title = "{Chern-Simons-matter conformal field theory on fuzzy sphere: Confinement transition of Kalmeyer-Laughlin chiral spin liquid}",
    eprint = "2507.19580",
    archivePrefix = "arXiv",
    primaryClass = "cond-mat.str-el",
    month = "7",
    year = "2025"
}

@misc{He:2025ong,
      title={Free real scalar CFT on fuzzy sphere: spectrum, algebra and wavefunction ansatz}, 
      author={Yin-Chen He},
      year={2025},
      eprint={2506.14904},
      archivePrefix={arXiv},
      primaryClass={hep-th},
      url={https://arxiv.org/abs/2506.14904}, 
}

@article{Taylor:2025odf,
    author = "Taylor, Joseph and Voinea, Cristian and Papi{\'c}, Zlatko and Fan, Ruihua",
    title = "{Conformal scalar field theory from Ising tricriticality on the fuzzy sphere}",
    eprint = "2506.22539",
    archivePrefix = "arXiv",
    primaryClass = "cond-mat.str-el",
    month = "6",
    year = "2025"
}

@article{ArguelloCruz:2025zuq,
    author = "Arguello Cruz, Erick and Klebanov, Igor R. and Tarnopolsky, Grigory and Xin, Yuan",
    title = "{Yang-Lee Quantum Criticality in Various Dimensions}",
    eprint = "2505.06369",
    archivePrefix = "arXiv",
    primaryClass = "hep-th",
    month = "5",
    year = "2025"
}

@article{Fan:2025bhc,
    author = "Fan, Ruihua and Dong, Junkai and Vishwanath, Ashvin",
    title = "{Simulating the non-unitary Yang-Lee conformal field theory on the fuzzy sphere}",
    eprint = "2505.06342",
    archivePrefix = "arXiv",
    primaryClass = "cond-mat.str-el",
    month = "5",
    year = "2025"
}

@article{EliasMiro:2025msj,
    author = "Elias Mir{\'o}, Joan and Delouche, Olivier",
    title = "{Flowing from the Ising Model on the Fuzzy Sphere to the 3D Lee-Yang CFT}",
    eprint = "2505.07655",
    archivePrefix = "arXiv",
    primaryClass = "hep-th",
    month = "5",
    year = "2025"
}

@article{Hu:2023xak,
    author = "Hu, Liangdong and He, Yin-Chen and Zhu, W.",
    title = "{Operator Product Expansion Coefficients of the 3D Ising Criticality via Quantum Fuzzy Spheres}",
    eprint = "2303.08844",
    archivePrefix = "arXiv",
    primaryClass = "cond-mat.stat-mech",
    doi = "10.1103/PhysRevLett.131.031601",
    journal = "Phys. Rev. Lett.",
    volume = "131",
    number = "3",
    pages = "031601",
    year = "2023"
}

@article{Han:2023yyb,
    author = "Han, Chao and Hu, Liangdong and Zhu, W. and He, Yin-Chen",
    title = "{Conformal four-point correlators of the three-dimensional Ising transition via the quantum fuzzy sphere}",
    eprint = "2306.04681",
    archivePrefix = "arXiv",
    primaryClass = "cond-mat.stat-mech",
    doi = "10.1103/PhysRevB.108.235123",
    journal = "Phys. Rev. B",
    volume = "108",
    number = "23",
    pages = "235123",
    year = "2023"
}

@article{Fardelli:2024qla,
    author = "Fardelli, Giulia and Fitzpatrick, A. Liam and Katz, Emanuel",
    title = "{Constructing the Infrared Conformal Generators on the Fuzzy Sphere}",
    eprint = "2409.02998",
    archivePrefix = "arXiv",
    primaryClass = "hep-th",
    doi = "10.21468/SciPostPhys.18.3.086",
    month = "9",
    year = "2024"
}

@article{Fan:2024vcz,
    author = "Fan, Ruihua",
    title = "{Note on explicit construction of conformal generators on the fuzzy sphere}",
    eprint = "2409.08257",
    archivePrefix = "arXiv",
    primaryClass = "hep-th",
    month = "9",
    year = "2024"
}

@article{Voinea:2024ryq,
    author = "Voinea, Cristian and Fan, Ruihua and Regnault, Nicolas and Papi{\'c}, Zlatko",
    title = "{Regularizing 3D Conformal Field Theories via Anyons on the Fuzzy Sphere}",
    eprint = "2411.15299",
    archivePrefix = "arXiv",
    primaryClass = "cond-mat.stat-mech",
    doi = "10.1103/bf4k-phl9",
    journal = "Phys. Rev. X",
    volume = "15",
    number = "3",
    pages = "031007",
    year = "2025"
}

@article{Hu:2023ghk,
    author = "Hu, Liangdong and He, Yin-Chen and Zhu, W.",
    title = "{Solving conformal defects in 3D conformal field theory using fuzzy sphere regularization [doi: 10.1038/s41467-024-47978-y]}",
    eprint = "2308.01903",
    archivePrefix = "arXiv",
    primaryClass = "cond-mat.stat-mech",
    doi = "10.1038/s41467-024-52959-2",
    journal = "Nature Commun.",
    volume = "15",
    number = "1",
    pages = "9013",
    year = "2024"
}

@article{Zhou:2023fqu,
    author = "Zhou, Zheng and Gaiotto, Davide and He, Yin-Chen and Zou, Yijian",
    title = "{The $g$-function and defect changing operators from wavefunction overlap on a fuzzy sphere}",
    eprint = "2401.00039",
    archivePrefix = "arXiv",
    primaryClass = "hep-th",
    doi = "10.21468/SciPostPhys.17.1.021",
    journal = "SciPost Phys.",
    volume = "17",
    number = "1",
    pages = "021",
    year = "2024"
}

@article{Cuomo:2024psk,
    author = "Cuomo, Gabriel and He, Yin-Chen and Komargodski, Zohar",
    title = "{Impurities with a cusp: general theory and 3d Ising}",
    eprint = "2406.10186",
    archivePrefix = "arXiv",
    primaryClass = "hep-th",
    doi = "10.1007/JHEP11(2024)061",
    journal = "JHEP",
    volume = "11",
    pages = "061",
    year = "2024"
}

@article{Zhou:2024dbt,
    author = "Zhou, Zheng and Zou, Yijian",
    title = "{Studying the 3d Ising surface CFTs on the fuzzy sphere}",
    eprint = "2407.15914",
    archivePrefix = "arXiv",
    primaryClass = "hep-th",
    doi = "10.21468/SciPostPhys.18.1.031",
    journal = "SciPost Phys.",
    volume = "18",
    number = "1",
    pages = "031",
    year = "2025"
}

@article{Dedushenko:2024nwi,
    author = "Dedushenko, Mykola",
    title = "{Ising BCFT from Fuzzy Hemisphere}",
    eprint = "2407.15948",
    archivePrefix = "arXiv",
    primaryClass = "hep-th",
    month = "7",
    year = "2024"
}

@article{Gao:2025vho,
    author = "Gao, Zhi-Qiang and Wang, Taige and Lee, Dung-Hai",
    title = "{Interacting Chern insulator transition on the sphere: revealing the Gross-Neveu-Yukawa criticality}",
    eprint = "2504.15338",
    archivePrefix = "arXiv",
    primaryClass = "cond-mat.str-el",
    month = "4",
    year = "2025"
}

@article{Hu:2024pen,
    author = "Hu, Liangdong and Zhu, W. and He, Yin-Chen",
    title = "{Entropic F function of three-dimensional Ising conformal field theory via fuzzy sphere regularization}",
    eprint = "2401.17362",
    archivePrefix = "arXiv",
    primaryClass = "hep-th",
    doi = "10.1103/PhysRevB.111.155151",
    journal = "Phys. Rev. B",
    volume = "111",
    number = "15",
    pages = "155151",
    year = "2025"
}

@article{Jafferis:2011zi,
    author = "Jafferis, Daniel L. and Klebanov, Igor R. and Pufu, Silviu S. and Safdi, Benjamin R.",
    title = "{Towards the F-Theorem: N=2 Field Theories on the Three-Sphere}",
    eprint = "1103.1181",
    archivePrefix = "arXiv",
    primaryClass = "hep-th",
    reportNumber = "PUPT-2366",
    doi = "10.1007/JHEP06(2011)102",
    journal = "JHEP",
    volume = "06",
    pages = "102",
    year = "2011"
}

@article{Closset:2012vg,
    author = "Closset, Cyril and Dumitrescu, Thomas T. and Festuccia, Guido and Komargodski, Zohar and Seiberg, Nathan",
    title = "{Contact Terms, Unitarity, and F-Maximization in Three-Dimensional Superconformal Theories}",
    eprint = "1205.4142",
    archivePrefix = "arXiv",
    primaryClass = "hep-th",
    reportNumber = "PUPT-2407, WIS-05-12-FEB-DPPA",
    doi = "10.1007/JHEP10(2012)053",
    journal = "JHEP",
    volume = "10",
    pages = "053",
    year = "2012"
}

@article{Klebanov:2011gs,
    author = "Klebanov, Igor R. and Pufu, Silviu S. and Safdi, Benjamin R.",
    title = "{F-Theorem without Supersymmetry}",
    eprint = "1105.4598",
    archivePrefix = "arXiv",
    primaryClass = "hep-th",
    reportNumber = "PUPT-2377",
    doi = "10.1007/JHEP10(2011)038",
    journal = "JHEP",
    volume = "10",
    pages = "038",
    year = "2011"
}

@article{Casini:2012ei,
    author = "Casini, H. and Huerta, Marina",
    title = "{On the RG running of the entanglement entropy of a circle}",
    eprint = "1202.5650",
    archivePrefix = "arXiv",
    primaryClass = "hep-th",
    doi = "10.1103/PhysRevD.85.125016",
    journal = "Phys. Rev. D",
    volume = "85",
    pages = "125016",
    year = "2012"
}

@article{Giombi:2014xxa,
    author = "Giombi, Simone and Klebanov, Igor R.",
    title = "{Interpolating between $a$ and $F$}",
    eprint = "1409.1937",
    archivePrefix = "arXiv",
    primaryClass = "hep-th",
    reportNumber = "PUPT-2472",
    doi = "10.1007/JHEP03(2015)117",
    journal = "JHEP",
    volume = "03",
    pages = "117",
    year = "2015"
}

@article{Nambu:1961tp,
    author = "Nambu, Yoichiro and Jona-Lasinio, G.",
    editor = "Eguchi, T.",
    title = "{Dynamical Model of Elementary Particles Based on an Analogy with Superconductivity. 1.}",
    doi = "10.1103/PhysRev.122.345",
    journal = "Phys. Rev.",
    volume = "122",
    pages = "345--358",
    year = "1961"
}

@article{Jack:2024sjr,
    author = "Jack, Ian and Osborn, Hugh and Steudtner, Tom",
    title = "{Explorations in scalar fermion theories: {\ensuremath{\beta}}-functions, supersymmetry and fixed points}",
    eprint = "2301.10903",
    archivePrefix = "arXiv",
    primaryClass = "hep-th",
    reportNumber = "DO-TH 22/06",
    doi = "10.1007/JHEP02(2024)038",
    journal = "JHEP",
    volume = "02",
    pages = "038",
    year = "2024"
}

@article{Lee:2006if,
    author = "Lee, Sung-Sik",
    title = "{Emergence of supersymmetry at a critical point of a lattice model}",
    eprint = "cond-mat/0611658",
    archivePrefix = "arXiv",
    doi = "10.1103/PhysRevB.76.075103",
    journal = "Phys. Rev. B",
    volume = "76",
    pages = "075103",
    year = "2007"
}

@article{Bobev:2015vsa,
    author = "Bobev, Nikolay and El-Showk, Sheer and Mazac, Dalimil and Paulos, Miguel F.",
    title = "{Bootstrapping the Three-Dimensional Supersymmetric Ising Model}",
    eprint = "1502.04124",
    archivePrefix = "arXiv",
    primaryClass = "hep-th",
    doi = "10.1103/PhysRevLett.115.051601",
    journal = "Phys. Rev. Lett.",
    volume = "115",
    number = "5",
    pages = "051601",
    year = "2015"
}

@article{Bobev:2015jxa,
    author = "Bobev, Nikolay and El-Showk, Sheer and Mazac, Dalimil and Paulos, Miguel F.",
    title = "{Bootstrapping SCFTs with Four Supercharges}",
    eprint = "1503.02081",
    archivePrefix = "arXiv",
    primaryClass = "hep-th",
    doi = "10.1007/JHEP08(2015)142",
    journal = "JHEP",
    volume = "08",
    pages = "142",
    year = "2015"
}

@article{Li:2017kck,
    author = "Li, Zhijin and Su, Ning",
    title = "{3D CFT Archipelago from Single Correlator Bootstrap}",
    eprint = "1706.06960",
    archivePrefix = "arXiv",
    primaryClass = "hep-th",
    doi = "10.1016/j.physletb.2019.134920",
    journal = "Phys. Lett. B",
    volume = "797",
    pages = "134920",
    year = "2019"
}

@article{Giombi:2021cnr,
    author = "Giombi, Simone and Helfenberger, Elizabeth and Khanchandani, Himanshu",
    title = "{Fermions in AdS and Gross-Neveu BCFT}",
    eprint = "2110.04268",
    archivePrefix = "arXiv",
    primaryClass = "hep-th",
    doi = "10.1007/JHEP07(2022)018",
    journal = "JHEP",
    volume = "07",
    pages = "018",
    year = "2022"
}

@article{Giombi:2022vnz,
    author = "Giombi, Simone and Helfenberger, Elizabeth and Khanchandani, Himanshu",
    title = "{Line defects in fermionic CFTs}",
    eprint = "2211.11073",
    archivePrefix = "arXiv",
    primaryClass = "hep-th",
    reportNumber = "PUPT-2644",
    doi = "10.1007/JHEP08(2023)224",
    journal = "JHEP",
    volume = "08",
    pages = "224",
    year = "2023"
}

@article{Barrat:2023ivo,
    author = "Barrat, Julien and Liendo, Pedro and van Vliet, Philine",
    title = "{Line defect correlators in fermionic CFTs}",
    eprint = "2304.13588",
    archivePrefix = "arXiv",
    primaryClass = "hep-th",
    reportNumber = "HU-EP-23/10-RTG, DESY-23-055",
    doi = "10.1007/JHEP05(2025)146",
    journal = "JHEP",
    volume = "05",
    pages = "146",
    year = "2025"
}

@article{Diatlyk:2024zkk,
    author = "Diatlyk, Oleksandr and Khanchandani, Himanshu and Popov, Fedor K. and Wang, Yifan",
    title = "{Defect fusion and Casimir energy in higher dimensions}",
    eprint = "2404.05815",
    archivePrefix = "arXiv",
    primaryClass = "hep-th",
    doi = "10.1007/JHEP09(2024)006",
    journal = "JHEP",
    volume = "09",
    pages = "006",
    year = "2024"
}

@article{Herzog:2022jlx,
    author = "Herzog, Christopher P. and Schaub, Vladimir",
    title = "{Fermions in boundary conformal field theory: crossing symmetry and E-expansion}",
    eprint = "2209.05511",
    archivePrefix = "arXiv",
    primaryClass = "hep-th",
    doi = "10.1007/JHEP02(2023)129",
    journal = "JHEP",
    volume = "02",
    pages = "129",
    year = "2023"
}

@article{Jiang:2025sfb,
    author = "Jiang, Huan and Ge, Yang and Jian, Shao-Kai",
    title = "{Boundary criticality for the Gross-Neveu-Yukawa models}",
    eprint = "2503.13247",
    archivePrefix = "arXiv",
    primaryClass = "cond-mat.str-el",
    month = "3",
    year = "2025"
}

@article{Giombi:2017rhm,
    author = "Giombi, S. and Kirilin, V. and Skvortsov, E.",
    title = "{Notes on Spinning Operators in Fermionic CFT}",
    eprint = "1701.06997",
    archivePrefix = "arXiv",
    primaryClass = "hep-th",
    reportNumber = "PUPT-2517, LMU-ASC-05-17",
    doi = "10.1007/JHEP05(2017)041",
    journal = "JHEP",
    volume = "05",
    pages = "041",
    year = "2017"
}

@article{Diab:2016spb,
    author = "Diab, Kenan and Fei, Lin and Giombi, Simone and Klebanov, Igor R. and Tarnopolsky, Grigory",
    title = "{On ${C}_{J}$ and ${C}_{T}$ in the Gross{\textendash}Neveu and O(N) models}",
    eprint = "1601.07198",
    archivePrefix = "arXiv",
    primaryClass = "hep-th",
    reportNumber = "PUPT-2496",
    doi = "10.1088/1751-8113/49/40/405402",
    journal = "J. Phys. A",
    volume = "49",
    number = "40",
    pages = "405402",
    year = "2016"
}

@article{Goykhman:2020tsk,
    author = "Goykhman, Mikhail and Sinha, Ritam",
    title = "{CFT data in the Gross-Neveu model}",
    eprint = "2011.07768",
    archivePrefix = "arXiv",
    primaryClass = "hep-th",
    doi = "10.1103/PhysRevD.103.125004",
    journal = "Phys. Rev. D",
    volume = "103",
    number = "12",
    pages = "125004",
    year = "2021"
}

@article{Pannell:2023tzc,
    author = "Pannell, William H. and Stergiou, Andreas",
    title = "{Scalar-fermion fixed points in the {\ensuremath{\varepsilon}} expansion}",
    eprint = "2305.14417",
    archivePrefix = "arXiv",
    primaryClass = "hep-th",
    doi = "10.1007/JHEP08(2023)128",
    journal = "JHEP",
    volume = "08",
    pages = "128",
    year = "2023"
}

@article{Giombi:2025evu,
    author = "Giombi, Simone and Pendse, Anurag",
    title = "{Line Defects with a Cusp in Fermionic CFTs}",
    eprint = "2511.08547",
    archivePrefix = "arXiv",
    primaryClass = "hep-th",
    month = "11",
    year = "2025"
}

@article{Liendo:2021wpo,
    author = "Liendo, Pedro and Rong, Junchen",
    title = "{Seeking SUSY fixed points in the 4 {\ensuremath{-}} {\ensuremath{\epsilon}} expansion}",
    eprint = "2107.14515",
    archivePrefix = "arXiv",
    primaryClass = "hep-th",
    reportNumber = "DESY-21-114",
    doi = "10.1007/JHEP12(2021)033",
    journal = "JHEP",
    volume = "12",
    pages = "033",
    year = "2021"
}

@article{Wen:2000,
  title = {Continuous Topological Phase Transitions between Clean Quantum Hall States},
  author = {Wen, Xiao-Gang},
  journal = {Phys. Rev. Lett.},
  volume = {84},
  issue = {17},
  pages = {3950--3953},
  numpages = {0},
  year = {2000},
  month = {Apr},
  publisher = {American Physical Society},
  doi = {10.1103/PhysRevLett.84.3950},
  url = {https://link.aps.org/doi/10.1103/PhysRevLett.84.3950}
}

@article{Zerf:2017zqi,
    author = "Zerf, Nikolai and Mihaila, Luminita N. and Marquard, Peter and Herbut, Igor F. and Scherer, Michael M.",
    title = "{Four-loop critical exponents for the Gross-Neveu-Yukawa models}",
    eprint = "1709.05057",
    archivePrefix = "arXiv",
    primaryClass = "hep-th",
    reportNumber = "DESY-17-133",
    doi = "10.1103/PhysRevD.96.096010",
    journal = "Phys. Rev. D",
    volume = "96",
    number = "9",
    pages = "096010",
    year = "2017"
}

@article{Ihrig:2018hho,
    author = "Ihrig, Bernhard and Mihaila, Luminita N. and Scherer, Michael M.",
    title = "{Critical behavior of Dirac fermions from perturbative renormalization}",
    eprint = "1806.04977",
    archivePrefix = "arXiv",
    primaryClass = "cond-mat.str-el",
    doi = "10.1103/PhysRevB.98.125109",
    journal = "Phys. Rev. B",
    volume = "98",
    number = "12",
    pages = "125109",
    year = "2018"
}

@article{Ponte:2012ru,
    author = "Ponte, Pedro and Lee, Sung-Sik",
    title = "{Emergence of supersymmetry on the surface of three dimensional topological insulators}",
    eprint = "1206.2340",
    archivePrefix = "arXiv",
    primaryClass = "cond-mat.str-el",
    doi = "10.1088/1367-2630/16/1/013044",
    journal = "New J. Phys.",
    volume = "16",
    number = "1",
    pages = "013044",
    year = "2014"
}

@article{Voinea:2025iun,
    author = "Voinea, Cristian and Zhu, Wei and Regnault, Nicolas and Papi{\'c}, Zlatko",
    title = "{Critical Majorana fermion at a topological quantum Hall bilayer transition}",
    eprint = "2509.08036",
    archivePrefix = "arXiv",
    primaryClass = "cond-mat.str-el",
    month = "9",
    year = "2025"
}

@article{Zhou:2025kng,
    author = "Zhou, Zheng and Gaiotto, Davide and He, Yin-Chen",
    title = "{Free Majorana Fermion Meets Gauged Ising Conformal Field Theory on the Fuzzy Sphere}",
    eprint = "2509.08038",
    archivePrefix = "arXiv",
    primaryClass = "hep-th",
    month = "9",
    year = "2025"
}

@article{Nakayama:2022svf,
    author = "Nakayama, Yu and Kikuchi, Ken",
    title = "{The fate of non-supersymmetric Gross-Neveu-Yukawa fixed point in two dimensions}",
    eprint = "2212.06342",
    archivePrefix = "arXiv",
    primaryClass = "hep-th",
    doi = "10.1007/JHEP03(2023)240",
    journal = "JHEP",
    volume = "03",
    pages = "240",
    year = "2023"
}

@misc{diagham,
  Note= {Diag{H}am, \url{https://www.nick-ux.org/diagham}}
}

@misc{FuzzifiED,
      title={{FuzzifiED -- Julia package for numerics on the fuzzy sphere}}, 
      author={Zheng Zhou},
      year={2025},
      eprint={2503.00100},
      archivePrefix={arXiv},
      primaryClass={cond-mat.str-el},
      url={https://arxiv.org/abs/2503.00100}, 
}

\clearpage 


\clearpage 
\pagebreak

\onecolumngrid
\begin{center}
\textbf{\large Supplementary Materials for ``Emergence of 3D Superconformal Ising Criticality on the Fuzzy Sphere" }\\[5pt]
\vspace{0.1cm}
\begin{quote}
{\small 
In this Supplementary Material, we present a detailed introduction to the construction of our fuzzy sphere model, discussing in detail its symmetries and the connections between certain field theory operators in the Gross-Neveu-Yukawa theory. Moreover, we provide details of the numerical method for locating the superconformal fixed point and the employed normalization scheme. 
} \\[20pt]
\end{quote}
\end{center}
\setcounter{equation}{0}
\setcounter{figure}{0}
\setcounter{table}{0}
\setcounter{page}{1}
\setcounter{section}{0}
\makeatletter
\renewcommand{\theequation}{S\arabic{equation}}
\renewcommand{\thefigure}{S\arabic{figure}}
\renewcommand{\thesection}{S\Roman{section}}
\renewcommand{\thepage}{S\arabic{page}}
\renewcommand{\thetable}{S\arabic{table}}

\vspace{0cm}

\section{A. Model}

We simulate the $N=1$ Gross-Neveu-Yukawa (GNY) model by coupling two distinct quantum critical systems on concentric fuzzy spheres. This construction leverages the state-operator correspondence inherent to radial quantization, enabling direct access to conformal data through energy spectrum analysis \cite{Cardy:1984epx,Cardy:1985lth}. Firstly, we consider a fuzzy sphere model realizing the 3D Ising CFT,  following the approach proposed in \cite{Zhu:2022gjc}. This Hamiltonian takes the form:
\begin{align}
		H_{\text{Ising}}(U_0,U_1,h) = &    (2s+1)^2 \int \mathbf{d}\Omega_f^a \mathbf{d}\Omega_f^b  \,  U(\Omega_f^{ab}) \left[n_f^0(\theta_a, \varphi_a)n_f^0(\theta_b, \varphi_b)- n_f^z(\theta_a, \varphi_a)n_f^z(\theta_b, \varphi_b) \right] 
		\nonumber \\
		&- h (2s+1)  \int  \mathbf{d} \Omega_f  \, n^x_f(\theta,\varphi) , \label{eq:hamIsing}
\end{align}
where $c_m = (c_{m\uparrow}, c_{m\downarrow})$ denote fermionic operators in the lowest Landau level (LLL) orbitals labelled by $m = -s, \dots, s$, \(n_f^i = \sum_m c_m^\dagger \sigma^i_f c_m\) is fermion density operator. $U(\Omega_f^{ab})$ is a short-range interaction potential, which can be parameterized by Haldane pseudopotentials $U_{0}$ and $U_1$. The field $h$ tunes the Ising transition from a ferromagnet to a paramagnet. The number of fermions is fixed at $N_e = 2s+1$ and $s$ is the monopole charge at the center of the sphere \cite{Haldane1983}. This model exhibits exact $\mathrm{SO(3)}$ rotational symmetry, $\mathbb{Z}_2$ Ising symmetry, and particle-hole symmetries. Crucially, at the optimal critical point ($h_c \approx 0.316$ for $U_0=0.475$, $U_1=0.1$), critical excitations of the neutral spin degree of freedom  match the 3D Ising operator spectrum with remarkable accuracy \cite{Zhu:2022gjc}. 

Second, we consider a bilayer quantum Hall system tuned to the Halperin(220) -- Pfaffian transition \cite{Voinea:2025iun}:
\begin{eqnarray}
	H_{\text{Majorana}} (V_0^{\textrm{intra}},V_0^{\textrm{inter}},t)&= &H_{\text{intra}} + H_{\text{inter}} + H_t
	\nonumber \\
	&=&  \int \mathbf{d}\Omega_b^a  \mathbf{d}\Omega_b^b  \,  \sum_{\sigma=\uparrow,\downarrow} V^{\textrm{intra}}(\Omega_b^{ab}) n_b^{\sigma}(\theta_a, \varphi_a)n_b^{\sigma}(\theta_b, \varphi_b) 
	\nonumber \\
	& & + \int \mathbf{d}\Omega_b^a \mathbf{d}\Omega_b^b  \,  \sum_{\sigma=\uparrow,\downarrow} V^{\textrm{inter}}(\Omega_b^{ab}) n_b^{\sigma}(\theta_a, \varphi_a)n_b^{-\sigma}(\theta_b, \varphi_b)+t\sum_{i=0}^{N_s-1} \left( b_{i,\uparrow}^\dagger b_{i,\downarrow} + b_{i,\downarrow}^\dagger b_{i,\uparrow}\right). \label{eq:hamMF}
\end{eqnarray}
This Hamiltonian contains an intra-layer ($H^{\text{intra}}$) repulsive interaction, an inter-layer ($H^{\text{inter}}$) repulsive interaction, and a transverse tunneling term $H_t$. $b_m = (b_{m\uparrow}, b_{m\downarrow})$ are bosonic operators in LLL orbitals ($m = -s, \dots, s$), \(n_b^{i=0,x,y,z} = \sum_m b^{\dagger}_m \sigma^i b_m\) is the boson density operator. 
This transition corresponds to condensation of quasiparticle-quasihole pairs between layers, breaking the symmetry down to $\mathbb{Z}_2$ and yielding two decoupled Majorana species, $\psi_o$ ad $\psi_e$ with opposite parity under layer exchange \cite{Wen:2000}. At the optimal critical point ($V_0^{\text{intra}} = 1.0$, $V_0^{\text{inter}} = 0.48$, $t \approx 0.58$), the operator spectrum contains integer-spin descendants in the even-particles sector and half-integer spins in the odd-particles sector, consistent with the free Majorana fermion CFT, where the lowest non-trivial scalar primary $\bar{\psi}\psi$ has $\Delta_{\bar{\psi}\psi}=2$ and fermion operator $\psi$ exhibits scaling dimension $\Delta_\psi = 1$.

Finally, we consider 
a Yukawa-like interaction between the Ising order parameter $\phi$ and the Majorana field $\psi$:
$H_{\text{coupling}} = g \phi \bar{\psi}\psi.$
This coupling preserves the combined $\mathrm{SO(3)}$ rotation and the time-reversal parity (which transforms as $\phi \to -\phi$ and $\bar{\psi}\psi \to - \bar{\psi}\psi$). 
Guided by this idea, we construct the coupled model via the following Hamiltonian, which was schematically depicted in Fig.~1(a):  
\begin{eqnarray}
	\label{eq:h_coupled}
	H (U_0,U_1,h,r,V_0^{\textrm{intra}},V_0^{\textrm{inter}},t,\{ \lambda \}) &=& H_{\text{Ising}} \otimes \mathbb{I}_{\text{Majorana}} + \mathbb{I}_{\text{Ising}} \otimes r H_{\text{Majorana}} 
	\nonumber \\
	& &+ \int \mathbf{d}\Omega_f  \mathbf{d}\Omega_b  \left[ \lambda^{z0}(\Omega_{fb})  n_f^z(\Omega_f)n_b^0(\Omega_b) +\lambda^{zx}(\Omega_{fb})n_f^z(\Omega_f) n_b^x(\Omega_b)  \right].
\end{eqnarray}
Here, $\lambda^{z0}(\Omega_{fb}),\lambda^{zx}(\Omega_{fb})$ describe the interactions between fermions and bosons. 
These local interactions are parameterized by the Haldane pseudopotentials $\{ \lambda_0^{z0},\lambda_1^{z0},\lambda_0^{zx},\lambda_1^{zx} \}$. We tune four parameters to reach the putative GNY fixed point, because there are three relevant scalar fields $\sigma,\sigma',\epsilon$ in the operator spectra (see below). 
The parameter $r$ matches the speed of light of the Majorana fermion and the Ising CFT.
In this paper, we fix the parameters of fermion layer (boson layer) to critical values that produces 3D Ising CFT (Majorana CFT).
When the coupled parameters were tunned to  \(\lambda = 0\), the system is equivalent to the tensor product of two independent CFTs (Ising CFT $\otimes$Majorana CFT). For certain \(\lambda = \lambda_c\), the RG drives the coupled system to flow to GNY fixed point which is expected to be captured by the 3D \(\mathcal{N} = 1\) supersymmetric Ising CFT \cite{Fei:2016sgs}.  

\subsection{A.1 Symmetries of the model}

The fermionic fuzzy sphere model, Eq.~\eqref{eq:hamIsing}, exhibits three essential symmetries. First, the $\mathrm{SO(3)}$ rotational symmetry acts on the orbital indices $m$, corresponding to Lorentz symmetry in the IR conformal field theory. Second, an Ising $\mathbb{Z}_2$ spin-flip symmetry defined by $c_m \to \sigma^x c_m$ persists throughout the phase diagram. Third, a particle-hole symmetry transformation $c_m \to i\sigma^y c^*_m$ maps directly to spacetime parity in the IR Ising CFT \cite{Zhu:2022gjc}. The order parameter $n_f^z \sim \sum_m c^\dagger_m \sigma^z c_m$ transforms as a scalar under $\mathrm{SO(3)}$ rotations while being odd under both $\mathbb{Z}_2$ and parity transformations.

The bosonic fuzzy sphere model, Eq.~\eqref{eq:hamMF}, displays a continuous quantum phase transition between the Halperin-$(220)$ state and Moore-Read Pfaffian state, tuned by interlayer tunneling amplitude $t$ and density-density interactions. This system similarly preserves $\mathrm{SO(3)}$ rotational symmetry inherited from the spherical Landau level projection, along with an intact layer-exchange $\mathbb{Z}_2$ symmetry. However, the realization of time-reversal parity presents subtleties: while emerging at criticality, this symmetry is explicitly violated in both gapped phases by the Majorana mass term $\bar{\psi}\psi$, which transforms oddly under parity. Crucially, this parity lacks direct microscopic representation and is instead identified through operator spectrum analysis at criticality \cite{Voinea:2025iun}.

The coupled Hamiltonian $H$, Eq.~\eqref{eq:h_coupled}, maintains $\mathrm{SO(3)}$ rotational symmetry, as both subsystems and their coupling term transform as scalars under spherical rotations. Consequently, all eigenstates possess well-defined angular momentum quantum numbers $L$ corresponding directly to the Lorentz spin quantum numbers in the emergent CFT.

Implementation of discrete symmetries requires careful consideration to involve the Yukawa coupling. The GNY model preserves exclusively a combined $\mathbb{Z}_2$ symmetry (time-reversal symmetry), defined through a simultaneous application of the fermionic spin-flip transformation $c_m \to \sigma^x c_m$ and the bosonic parity transformation $\psi \to \gamma^0 \psi$ (the latter lacking a direct representation on the fuzzy sphere, see the discussion above). Neither symmetry remains individually invariant under the coupling interaction.

Moreover, for model calculations, local microscopic operators admit expansions in terms of corresponding CFT fields. At the Ising transition point in the fermionic fuzzy sphere model Eq. \eqref{eq:hamIsing}, the order parameter expands as $n^z_f = g_1 \phi + g_2 \phi^3 + \text{descendants} + \cdots$, where $\phi$ denotes the leading $\mathbb{Z}_2$-odd scalar primary operator with $\Delta_\phi \approx 0.518$ (note $\phi^3 \propto \Box \phi$ at criticality due to equations of motion, a relation broken by Yukawa interactions). In the bosonic fuzzy sphere model, both density operators $n^{0}_b$ and $n^{x}_b$ lack definite parity and expand as $n^{0/x}_b = g_3^{0/x} \mathbb{I} + g_4^{0/x} \bar{\psi}\psi + \text{descendants} + \cdots$, where $\mathbb{I}$ is the identity operator in the Majorana fermion CFT.
Consquently, when microscopic parameters $(U_0,U_1,h,V_0^{\text{intra}},V_0^{\text{inter}},t)$ are tuned to critical values, the boson-fermion interaction in Eq. \eqref{eq:h_coupled} should expand as:
\begin{equation}
	\begin{aligned}
       \lambda^{z0}(\Omega_{fb})  n_f^z(\Omega_f)n_b^0(\Omega_b) +\lambda^{zx}(\Omega_{fb})n_f^z(\Omega_f) n_b^x(\Omega_b) 
       \sim 
        (g_1\phi + g_2\phi^3)\left[g_3(\{\lambda\}) \mathbb{I} + g_4(\{\lambda\}) \overline{\psi}\psi\right] 
		+ \text{Higher order terms},
	\end{aligned}
    \label{eq:effe-Lag}
\end{equation}
where the coefficients $g_3(\{\lambda\})$ and $g_4(\{\lambda\})$ depend on the Haldane pseudopotentials $\lambda^{z0}$ and $\lambda^{zx}$. Realization of $\mathcal{N}=1$ supersymmetry requires fine-tuning to achieve $g_3(\{\lambda_c\}) = 0$. Under this condition, the identity operator $\mathbb{I}$ vanishes, reducing the leading contribution to the Yukawa interaction $g_1g_4 \phi\bar{\psi}\psi$ and restoring parity symmetry throughout the coupled system. These specific conditions drive the initially decoupled Ising and Majorana CFTs toward the $N=1$ GNY fixed point with emergent supersymmetry.

Our model exhibits two additional  symmetries under which the energy spectrum remains invariant:

1. $\{\lambda\} \to \{-\lambda\}$:
   This is consistent with the effective Lagrangian in Eq.~\eqref{eq:effe-Lag}, since all additional perturbations generated by this term are invariant under this reflection transformation. (The symmetry of the Yukawa interaction can also be directly seen from Fig.~\ref{fig:combined}(c).)

2. $\{ t, \lambda^{z0}, \lambda^{zx} \} \to \{ -t, -\lambda^{z0}, \lambda^{zx} \}$:
   In the original bosonic model, flipping the inter-layer boson tunneling amplitude ($t \to -t$) leaves the full spectrum unchanged. However, in the realized Majorana fermion CFT, the local operators $n^0$ and $n^x$ have different operator-content expansion coefficient. As a result, once the fermion–boson interaction is introduced, the transformations of the two couplings must differ by a relative minus sign in order for the full Hamiltonian spectrum to remain invariant.

\subsection{A.2 Operator correspondence}

We now analyze the field theoretical content and its connection to microscopic operators. The most relevant scalar operator $\sigma$ exhibits distinct symmetry properties: it is $\mathbb{Z}_2$-odd in the Ising CFT while becoming parity-odd at the GNY fixed point. Crucially, since our fuzzy sphere model lacks manifest parity symmetry throughout most parameter space regions, generic points suffer $\phi$-induced perturbations. Given its relatively small scaling dimension, parity violation induces significant deviations from the GNY fixed point.

The $\mathbb{Z}_2$-even singlet operator $\epsilon$ warrants separate consideration. In pure Ising systems, this operator drives the paramagnet-to-ferromagnet transition, while in GNY theory it simultaneously gaps both bosonic and fermionic sectors, distinct from Ising universality class. Pretuning $U_0$, $U_1$, and $h$ to critical values eliminates this relevant perturbation. After establishing the GNY fixed point, controlled variation of the transverse field strength $h$ enables observation of phase transition belonging to GNY universality class.

Within the Ising critical manifold, a continuous line of second-order phase transitions emerges under variation of $U_0$ and $h$. While all points flow to the 3D Ising CFT in the IR, they possess distinct UV deformations through the $\phi^4$ coupling, corresponding to the subleading $\mathbb{Z}_2$-even operator $\epsilon'$. During the flow to the GNY fixed point, scalar operators $\phi^4$ and $\phi\Box\phi$ recombine into novel primary fields ($\epsilon'$ and $\epsilon''$). Although explicit supersymmetry requires $g^2 = 2\lambda$, both operators remain irrelevant at the GNY fixed point ($\Delta_{\epsilon'} \approx 3.89$, $\Delta_{\epsilon''} \approx 4.38$) \cite{Atanasov:2022bpi}, ensuring the emergence of criticality and supersymmetry. This allows arbitrary selection of phase transition points along the critical line. Recent work identifies the free Gaussian CFT as an Ising tricritical point where the Ising transition meets a first-order phase transition \cite{Taylor:2025odf}. Coupling such systems with Majorana fields could enable the searching of the tricritical GNY$^*$ CFT discussed in \cite{Fei:2016sgs,Iliesiu:2017nrv} (see also Fig.~\ref{fig:combined}).

The parameter $r$ in Eq.~\eqref{eq:h_coupled} controls the relative speed of light between Ising and Majorana sectors. While this anisotropy should ideally be normalized to unity in the sense of field theory, renormalization group analysis suggests its irrelevance \cite{Grover:2013rc}. Our numerical simulations—though limited by moderate system sizes—demonstrate robust emergence of the GNY fixed point insensitive to $r$, consistent with RG predictions. We set $r=1$ in the following discussion.

Finally, we posit that at critical coupling $\lambda_c$, the boson-fermion interacting term corresponds directly to perturbation under the Yukawa coupling operator. Adiabatic tuning of $\lambda$ from $0$ to $\lambda_c$ approximately traverses the RG flow connecting the ultraviolet fixed point (decoupled Ising $\otimes$ MF CFT) and the infrared GNY fixed point. Crucially, the Yukawa coupling evolves from relevant to irrelevant during this flow, enabling detailed study of scaling operator trajectories along the RG, which has been discussed in tha main text.

Table~\ref{tab:theory} lists the low-lying primaries of the 3D Ising SCFT. The theoretical values are estimated from conformal bootstrap \cite{Rong:2018okz,Atanasov:2018kqw,Atanasov:2022bpi,Erramilli:2022kgp}. The results from the fuzzy sphere will be presented in Sec.~C below. 

\begin{table}[h]
	\centering
    \caption{Operator content of the 3D Ising SCFT. $L (P)$ represents Lorentz spin (time-reversal parity).  The confomal bootstrap (CB) data is from Ref. \cite{Rong:2018okz,Atanasov:2018kqw,Atanasov:2022bpi,Erramilli:2022kgp}.
    }
	\begin{tabular}{c|c|c|c|c|c}
		\hline
		\hline 
		Operator & $L$ & $P$ & $\Delta$ (CB) & \text{operator content in GNY model} & \text{Physical meaning} \\
        \hline
		$\sigma$ & 0 & -1 & 0.584 & $\sim \phi$ & magnetic operator \\ \hline 
        $\sigma'$ & 0 & -1 & 2.887 & $\sim \phi^3$ & subleading magnetic operator \\ 
		 \hline   
         $\epsilon$ & 0 & 1 & 1.584 & $\sim \phi^2$ & mass operator \\
		\hline
        $\epsilon'$ & 0 & 1 & 3.887 & $\sim \phi^4 + \phi\square \phi $ & subleading mass operator \\
        \hline 
        $\epsilon''$ & 0 & 1 & 4.38 & $\sim \phi^4 + \phi\square \phi $ & \\
        \hline 
        $\psi$ & 1/2 & 1 & 1.084 & $\sim \psi$ & Majorana fermion operator \\
        \hline 
        $\psi'$ & 1/2 & 1 & 3.387 & $\sim \phi^2 \psi$ & subleading fermion operator \\
        \hline 
        $G_{\mu}$ & 3/2 & -1 & 2.500 & $\sim (\partial_{\nu} \partial_{\mu}\psi) \partial^{\mu} \phi$ & supercurrent \\
        \hline 
        $T_{\mu\nu}$ & 2 & 1 & 3.000 & & energy-momentum tensor \\
        \hline
        $T_{\mu\nu}'$ & 2 & 1 & 3.4 & & \\
        \hline 
        \hline
	\end{tabular}
	\label{tab:theory}
\end{table}

\section{B. Determining the supersymmetric fixed points} 

To identify the supersymmetric fixed point in our model, we employ exact diagonalization (ED) on the coupled fermion-boson system, Eq.~\eqref{eq:h_coupled}, defined on the fuzzy sphere. The system size is governed by the number of Landau orbitals, related to the monopole flux \(s\) via \(N_{o} = 2s + 1\) (note that the number of bosons $N_b=N_{o}+1$ and the number of fermions $N_f=N_{o}$ on sphere). The many-body Hilbert space grows fast as for each $N_b=N_f+1$, we have two layers of bosonic LLLs and two layers of fermionic LLLs (in total, we have $N_b$ bosons and $N_f$ fermions to fill in $2N_{o}$ bosonic orbitals and $2N_{o}$ fermionic orbitals.). Therefore, we concentrate on $s=3$ $(N_b=8$, $N_f=7)$ for integer spin sectors and $s=5/2$ $(N_b=7$, $N_f=6)$ for half-integer spin sectors in the following discussion. Crucially, radial quantization on \(\mathbb{S}^2 \times \mathbb{R}\) maps eigenstates of $H$ to scaling operators, i.e., $\delta E = v\Delta / R$, allowing detection of operator scaling dimensions ($\{\Delta\}$) and Lorentz spin quantum numbers via the energy spectrum ($\{E_n\}$).

\subsection{B.1 Integer momentum sector}

At the maximum numerically-accessible system size \(s=3\) , we numerically diagonalize the Hamiltonian and compute following energy gaps between specific eigenstates:
\begin{equation}
	\left\{
	\begin{aligned}
		& \delta E_{\epsilon} = E_{\epsilon} - E_{\sigma}, \delta E_{\partial \sigma} = E_{\partial \sigma} - E_{\sigma}, 
		\\
		& \delta E_{\Box \sigma} = E_{\Box \sigma} - E_{\sigma},  \delta E_{\partial\epsilon} = E_{\partial\epsilon} - E_{\sigma}, 
		\\
		& \delta E_{\partial^2\sigma} = E_{\partial^2\sigma} - E_{\sigma}, \delta E_{T} = E_{T} - E_{\text{gs}}, 
	\end{aligned}
	\right\}
\end{equation}
At the supersymmetric fixed point, these gaps should be proportional to the scaling dimension differences in the $\mathcal{N}=1$ supersymmetric Ising CFT. Due to the state-operator correspondence on $\mathbb{S}^2 \times \mathbb{R}$, the expected gaps should be proportional to $\{\Delta_\epsilon - \Delta_\sigma, \Delta_{\partial \sigma} - \Delta_\sigma, \Delta_{\Box \sigma} - \Delta_\sigma, \Delta_{\partial\epsilon} - \Delta_\sigma, \Delta_{\partial^2\sigma} - \Delta_\sigma, \Delta_T\} = \{1, 1, 2, 2, 2, 3\}$. 

For each parameter $\{\lambda\} = \{\lambda^{z0}_0, \lambda^{z0}_1, \lambda^{zx}_0, \lambda^{zx}_1\}$, we numerically compute the low-lying spectrum and minimize a cost function that quantifies deviations from these universal ratios:
\begin{equation}
	\begin{aligned}
		Q_1(\lambda, \alpha) = &\left| \alpha \delta E_{\epsilon} - 1 \right| + \left| \alpha \delta E_{\partial \sigma} - 1 \right| + \left| \alpha \delta E_{\Box \sigma} - 2 \right| 
		+ \left| \alpha \delta E_{\partial\epsilon} - 2 \right| + \left| \alpha \delta E_{\partial^2\sigma} - 2 \right| + \left| \alpha \delta E_{T} - 3 \right|,
	\end{aligned}
\end{equation}
where \(\alpha\) is optimized numerically to minimize $Q_1$ for fixed \(\lambda\). Parameter space scanning reveals a minimum at $\{\lambda_c^{s=3}\} = \{\lambda^{z0}_0 = 0.09, \lambda^{z0}_1 = 0.09, \lambda^{zx}_0 = 0.43, \lambda^{zx}_1 = -0.17\}$. Although the cost function $Q_1$ does not require any conformal data as input, in order to determine the location of the critical point more accurately, we can define it as the discrepancy between the ratio of energy gaps and the numerical bootstrap results. We find that the energy spectrum at the optimal points obtained through different optimization methods remains almost unchanged, which demonstrates the effectiveness of this method for locating the critical point and the robustness of the emergent superconformal symmetry. The obtained bosonic operator spectrum at the optimal point is presented in Table~\ref{tab:numerical} and \ref{tab:numerical2}.

\begin{figure}
	\centering
		\includegraphics[width=0.4\textwidth]{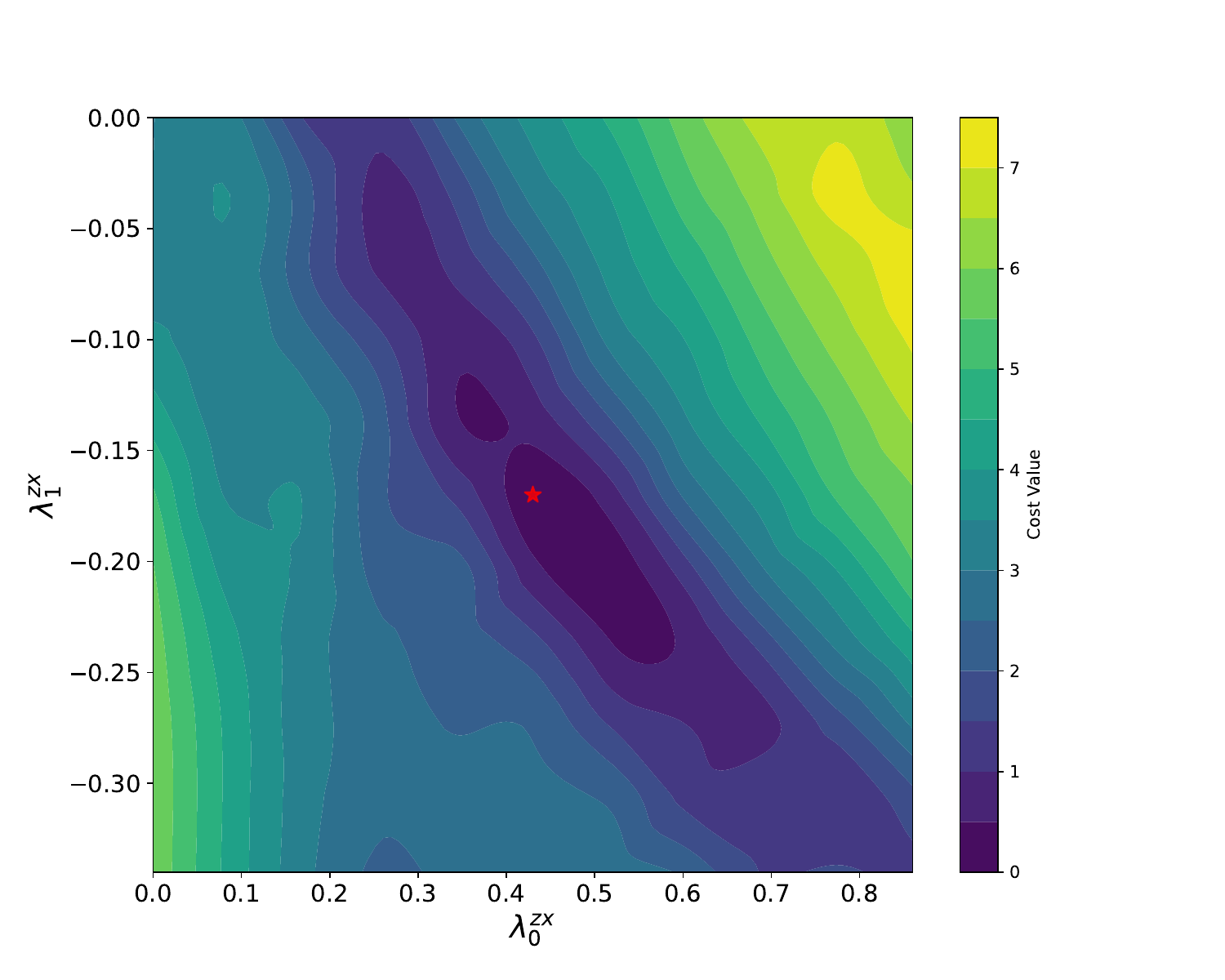}
        \includegraphics[width=0.4\textwidth]{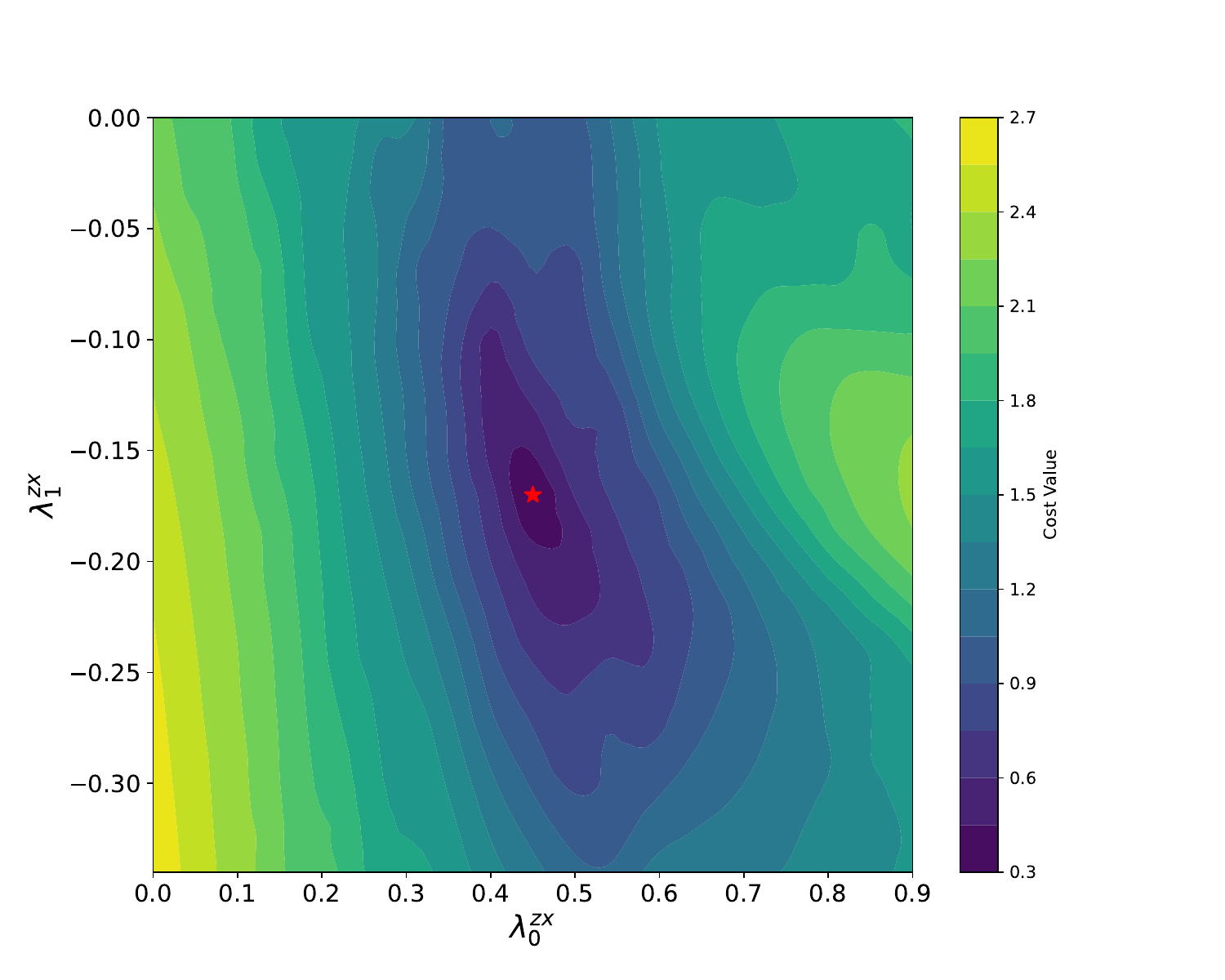}
	\caption{ The distribution of minimized cost function within the $\lambda^{zx}_0$-$\lambda^{zx}_1$ plane with the other two parameters fixed at critical values. Left for the integer momentum sector ($s=3$) and right for the half-integer momentum sector ($s=5/2$) with identified critical points marked by red star.  }
    \label{fig:cost_function}
\end{figure}

\begin{figure}[!htp]
	\centering
	\includegraphics[width=0.4\textwidth]{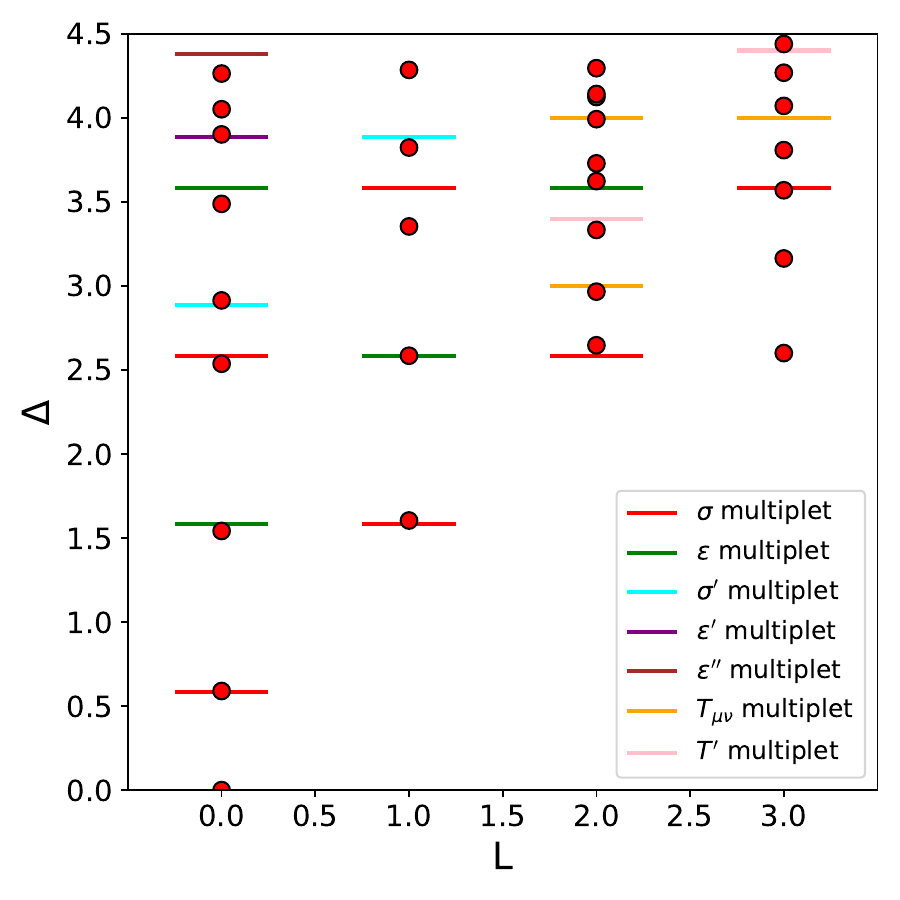}
	\includegraphics[width=0.4\textwidth]{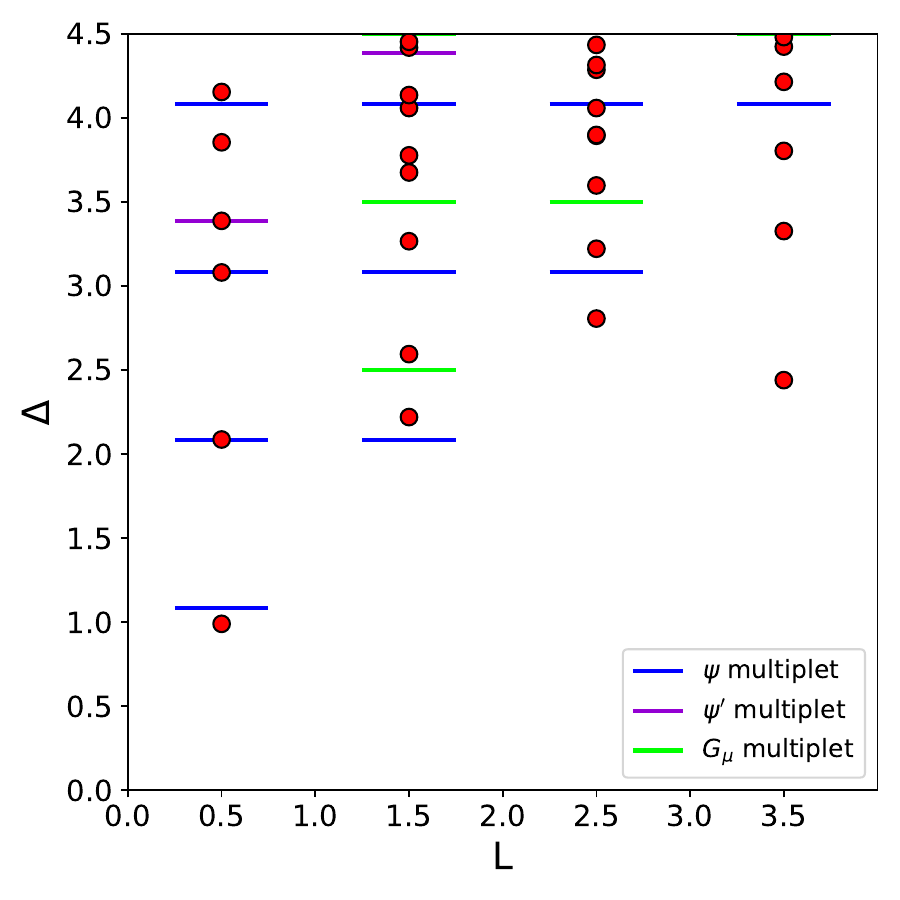}
	\caption{ Raw data of low-lying operator spectra for 3D super-Ising CFT.  (a) $Z_2$-even sector with $s= 3$ (8 bosons and 7 fermions) at $\lambda=\lambda_c^{s=3}$, containing integer angular momentum states. The whole spectrum is rescaled by minimizing the associated cost function. 
	(b) $Z_2$-even sector with $s=5/2$ (7 bosons and 6 fermions), containing half-integer angular momentum states. 
	Both panels show the complete energy spectra for levels with $\Delta \leq 4.5$, $L \leq 3.5$. The expected CFT operators are labeled, with their scaling dimensions computed from conformal bootstrap shown by short lines \cite{Rong:2018okz,Atanasov:2018kqw,Atanasov:2022bpi,Erramilli:2022kgp}. The rounded circles denotes numerical results calculated from our  Hamiltonian \eqref{eq:h_coupled}.
   }
	\label{fig:spectrum-c8}
\end{figure}

The left panel of Fig.~\ref{fig:cost_function} shows the distribution of the cost function \(Q_1\) in the \(\lambda^{zx}_0\)-\(\lambda^{zx}_1\) plane, with \(\lambda^{z0}_0\) and \(\lambda^{z0}_1\) fixed at critical values, exhibiting a pronounced minimum at \(\lambda_c\) that identifies a candidate SUSY fixed point. The obtained energy spectrum is shown in Fig.~\ref{fig:spectrum-c8}. We may also compute the half-integer momentum sector at this point. As shown in Fig.~\ref{fig:spectrum-c8}, the low-lying spectrum within $L=1/2$ sector exhibits a remarkable agreement with bootstrap prediction, while their descendants suffer from larger deviations due to finite-size effects. Thus, we search for the critical point within the half-momentum sector using the method presented in the next subsection.

\begin{figure}
	\centering
		\includegraphics[width=0.5\textwidth]{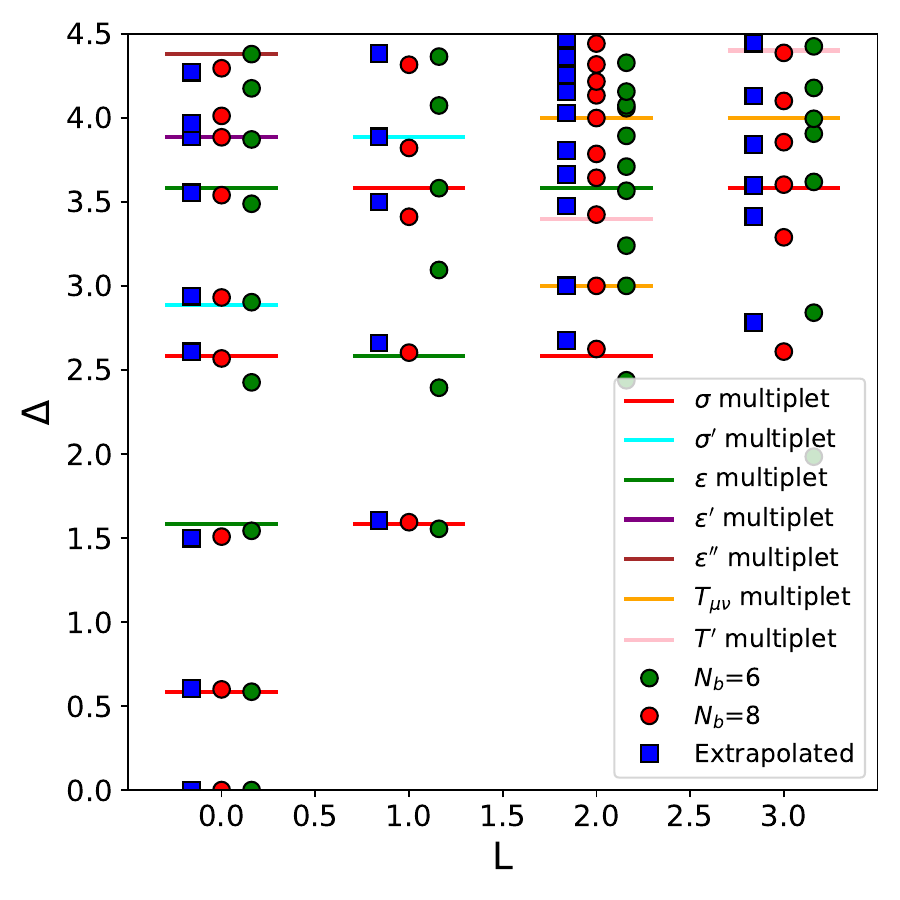}
	\caption{In this figure, the rescaled energy spectrum is represented by green (for $s=2$) and red (for $s=3$) circles, obtained by setting $\Delta_T=3$, using the same Hamiltonian parameters. The scaling dimension in the thermodynamic limit, derived from a simple polynomial extrapolation, is marked by blue squares. Predictions from numerical bootstrap are indicated by lines of corresponding colors. It can be observed that although some low-lying excited states deviate slightly from the bootstrap results under this treatment, higher excited states exhibit closer agreement with the theoretical values after extrapolation, better conforming to the tower structure expected from conformal symmetry.}
    \label{fig:extrapolated}
\end{figure}

Performing a direct finite-size scaling analysis on our numerical results is challenging for two main reasons. First, due to computational limitations, we are only able to simulate up to $s=3$ in the integer momentum sector. For $s \leq 1$, the errors are excessively large, so we focus exclusively on $s=2$ and $3$ for bosonic fields. Second, in the 3D super-Ising CFT, the scaling dimensions of the first and second irrelevant scalar operators are relatively close, making it necessary to account for corrections from both. Given these constraints, we adopt a simplified extrapolation procedure to process the data. Specifically, at $\lambda_c^{s=3}$, we compute the low-lying excitation spectra for 
$s=2$ and $3$ using ED, and rescale the entire spectrum by fixing the scaling dimension of the energy-momentum tensor to 3. We then perform a polynomial fit to extrapolate the predictions for the whole spectrum in the thermodynamic limit according to:
\begin{equation}
\Delta(N_{o}) = \Delta(\infty) + a/N_{o}^{\Delta_{\text{approx}}},
\end{equation}
where $\Delta_{\text{approx}}$ is optimized to minimize the cost function $Q_1$. The extrapolated spectrum obtained through this method, along with the original finite-size spectra, is presented in Fig.~\ref{fig:extrapolated}. It can be observed that the integer-spacing structure of some highly excited states becomes more pronounced after this treatment.

\begin{figure}
	\centering
		\includegraphics[width=0.4\textwidth]{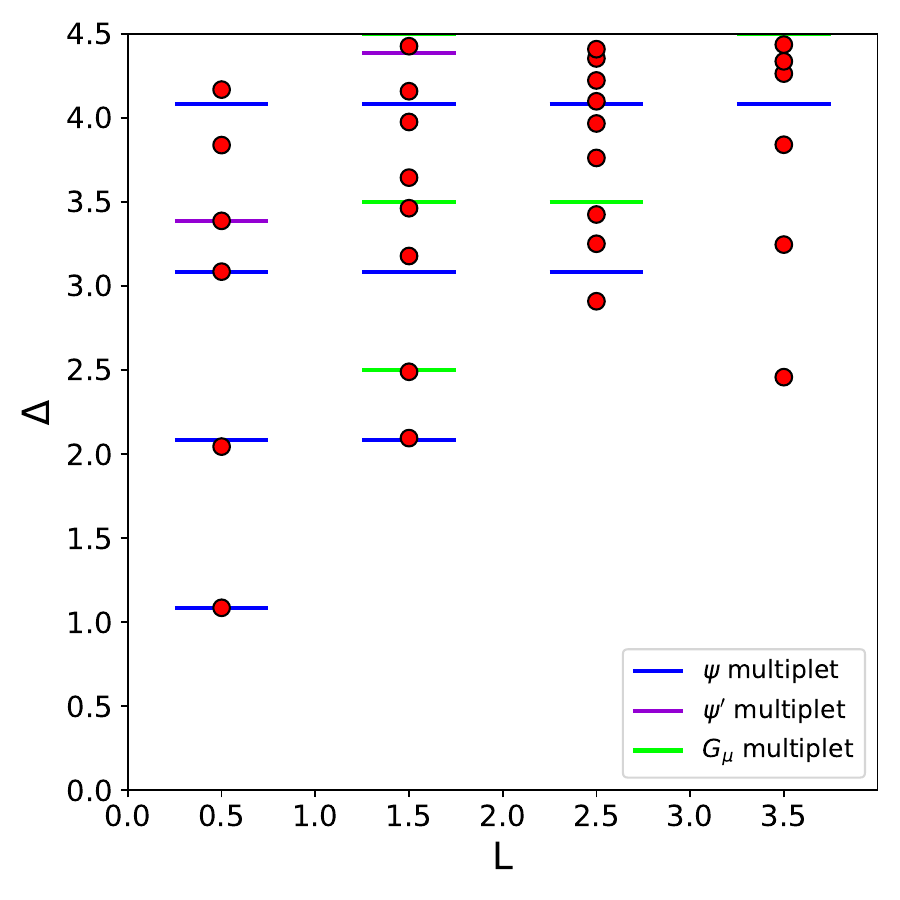}
	\caption{Raw data of low-lying spectrum computed at $\lambda=\lambda_c^{s=5/2}$ containing half-integer spin sectors. The derivation of spectra in the larger angular momentum sectors is from the finite-size effect (see discussion in the main text). }
    \label{fig:spectrum-c7}
\end{figure}

\subsection{B.2 Half-integer momentum sector}

For the half-integer momentum sector, we consider the system with an odd number of bosons, specifically $s=5/2$ \((N_b = 7,\) $N_f=6)$. In this case, all Lorentz spin quantum numbers are half-integers. As a result, the lowest energy state does not correspond to the vacuum state of the CFT, but rather to the fermionic field \(\psi\). This necessitates a different normalization procedure when identifying the conformal tower.

Moreover, due to the relatively small system sizes accessible to numerical exact diagonalization, the energy spectrum computed directly at the critical point determined from the \(s=3\) sector still shows certain deviations from the expected supersymmetric CFT values when \(s=5/2\). Therefore, we perform a renewed parameter scan in the vicinity of that point to identify the optimal set of couplings that best reproduces the conformal tower structure in the half-integer sector.

Specifically, for each parameter \(\{\lambda\}\), we employ ED to compute the energy spectrum of states corresponding to the following operators:
\begin{equation}
	\left\{
	\begin{aligned}
		 E_{\psi}, E_{\partial \psi}, E_{\sigma \psi}, E_{\partial^2 \psi}, E_{\Box \psi}, E_{\partial (\sigma \psi)}, E_{\psi'}, E_{G_{\mu}}
	\end{aligned}
	\right\}
\end{equation}
Due to radial quantization, at the supersymmetric fixed point these energies should be proportional to the corresponding scaling dimensions, i.e.,
\begin{equation}
	E_i = \alpha \Delta_i + \beta,
\end{equation}
up to a global rescaling factor \(\alpha\) and a constant shift \(\beta\).

Thus, for each parameter set \(\{\lambda\}\), we optimize over \(\alpha\) and \(\beta\) to minimize the following cost function:
\begin{equation}
	\begin{aligned}
		Q_2(\lambda, \alpha, \beta) = &\left| \alpha E_{\psi}+\beta - \Delta_\psi \right| +\left| \alpha E_{\partial\psi}+\beta - (\Delta_\psi+1) \right|
		+ \left| \alpha E_{\sigma\psi}+\beta - (\Delta_\psi+1) \right| 
        +\left| \alpha E_{\partial^2 \psi}+\beta - (\Delta_\psi+2) \right|
        \\ 
		&
		+ \left| \alpha E_{\Box\psi}+\beta - (\Delta_\psi+2) \right| +\left| \alpha E_{\partial (\sigma\psi)}+\beta - (\Delta_\psi+2) \right|
        +\left| \alpha E_{\psi'}+\beta - \Delta_\psi' \right|
	\end{aligned}
\end{equation}
where the expected scaling dimension differences are taken from the \(\mathcal{N}=1\) SUSY Ising CFT.

By scanning the parameter space, we find that the cost function \(Q_2\) is minimized at \(\lambda_c^{s=5/2} = \{\lambda^{z0}_0 = 0.15, \lambda^{z0}_1 = 0.06, \lambda^{zx}_0 = 0.45, \lambda^{zx}_1 = -0.17\}\). We identify this point as the critical point for the half-integer momentum sector, not very far from $\lambda_c^{s=3}$. The rescaled fermionic operator spectrum at the optimal point is presented in Tab.~\ref{tab:numerical} and \ref{tab:numerical2}.

In Fig.~\ref{fig:cost_function}, we also show the variation of \(Q_2\) in the \(\lambda^{zx}_0\)–\(\lambda^{zx}_1\) plane while keeping \(\lambda^{z0}_0\) and \(\lambda^{z0}_1\) fixed at their critical values. A clear minimum is observed at \(\lambda_c^{s=5/2}\), confirming the location of the SUSY fixed point in this sector. The resulting energy spectrum for the half-integer sector is presented in Fig.~\ref{fig:spectrum-c7}.
One may notice that the spectra in the larger angular momentum sector deviate more obviously from the expected values. We deduce there are two possible reasons: 1) The larger angular momentum sectors  suffer from larger finite-size effect. In other words, the state with relatively large angular momentum has a large expansion in spatial space. So, larger system sizes are generally required to capture these high-momentum states.
2) The magneto-rotor mode may appear in the high-momentum sectors, which does not belong to the CFT tower (see Ref. \cite{Voinea:2024ryq} for details). 

\section{C. Operator spectrum content}

In this section we explain in detail how the raw energy spectra are rescaled into scaling dimensions of the emergent CFT. Since different momentum sectors correspond to different system sizes, two distinct rescaling schemes are required. We present both procedures and summarize how they are used to obtain the spectra shown in Tabs.~\ref{tab:numerical} and \ref{tab:numerical2}.

For integer $L$ (the $s=3$ sector), the lowest-energy state corresponds to the identity operator with scaling dimension zero. Consequently, only a single global normalization factor~$\alpha$ is needed to relate the finite-size energy differences $\delta E_i$ to CFT scaling dimensions.

We adopt two distinct normalization schemes:

(1) For each point in the microscopic parameter space $(\lambda)$, we determine the optimal scaling factor $\alpha$ by minimizing the cost function $Q_1$. The function $Q_1$ measures the deviation between the rescaled gaps $\alpha \,\delta E_i$ and the theoretically predicted scaling dimensions of the corresponding SCFT operators. The optimal spectra obtained with this procedure are displayed in Tab.~\ref{tab:numerical}.

(2) An alternative scheme is to directly normalize the spectrum by imposing that the stress-tensor operator has scaling dimension~$3$. After implementing this normalization, we compare the resulting normalized energies with SCFT predictions and compute the corresponding cost function $Q_1$. The optimal spectra obtained with this second scheme are listed in Tab.~\ref{tab:numerical2}.

For half-integer $L$ (the $s=5/2$ sector), the situation is more subtle because the lowest energetic state in this sector corresponds to a nontrivial operator (the fermionic primary). As a result, in addition to an overall normalization factor~$\alpha$, a constant shift~$\beta$ is also required.

Again, we implement two rescaling procedures:

(1) For each set of microscopic couplings $\lambda$, we shift and rescale all energy levels according to 
\[E_i \to \alpha\, E_i + \beta ,\]
and simultaneously optimize $\alpha$ and $\beta$ to minimize the cost function $Q_2$. The function~$Q_2$ compares the resulting spectrum with numerical bootstrap predictions for the scaling dimensions of operators in the half-integer momentum sector. In this way, both the optimal spectrum and its normalization are determined dynamically. The resulting scaling dimensions are included in Tab.~\ref{tab:numerical}.

(2) A second normalization strategy is obtained by directly fixing the scaling dimensions of selected benchmark states. For example, we may normalize the energy difference $E_{\partial \psi}-E_{\psi}$ 
to unity, and simultaneously shift the candidate supercurrent operator $G_\mu$ (the second state in the $L=\tfrac32$ sector) to have scaling dimension $5/2$. These conditions determine the normalization parameters $(\alpha, \beta)$ for any $\lambda$. The half-integer sector spectra extracted using this scheme are presented in Tab.~\ref{tab:numerical2}.

These two complementary rescaling schemes allow us to verify the consistency of the extracted spectra and to compare the numerical data with SCFT predictions across the full parameter space. And one could find robust emergence of superconformal invariance under different normalization schemes.

\begin{table*}[htbp]
	\centering
	\caption{Comparison between the extracted scaling dimensions of low-lying primaries within the fuzzy sphere calculation and numerical bootstrap computation \cite{Iliesiu:2015qra,Iliesiu:2017nrv,Rong:2018okz,Atanasov:2018kqw,Atanasov:2022bpi,Erramilli:2022kgp,Mitchell:2024hix}. The original spectrum is rescaled by minimization of the corresponding cost functions.}
	\begin{tabular}{c|ccccc|ccc|cc}
		\hline
		\hline 
		Operator & $\sigma$ & $\epsilon$ & $\sigma'$ & $\epsilon'$ & $\epsilon''$ & $\psi$ & $\psi'$ & $G_\mu$ & $T_{\mu\nu}$ & $T'$ \\
		$L$ & 0 & 0 & 0 & 0 & 0 & 1/2 & 1/2 & 3/2 & 2 & 2 \\ \hline 
		$\Delta_{\text{CB}}$ & 0.584 & 1.584 & 2.887 & 3.887 & 4.38 & 1.084 & 3.387 & 2.5 & 3.0  & 3.4    \\
		$\Delta_{\text{FS}}$ & 0.589 & 1.542 & 2.913 & 3.901 & 4.051 & 1.084 & 3.387 & 2.489 & 2.966 & 3.333    \\ 
		Error & 0.87$\%$ & 2.69$\%$ & 0.92$\%$ & 0.37$\%$ & 7.52$\%$ & 0.00$\%$  & 0.01$\%$ & 0.44$\%$ & 1.15$\%$  & 1.98$\%$ \\ 
		\hline
	\end{tabular}
	\label{tab:numerical}
\end{table*}

\begin{table*}[htbp]
	\centering

    \caption{Comparison between the extracted scaling dimensions of low-lying primaries within the fuzzy sphere calculation and numerical bootstrap computation \cite{Iliesiu:2015qra,Iliesiu:2017nrv,Rong:2018okz,Atanasov:2018kqw,Atanasov:2022bpi,Erramilli:2022kgp,Mitchell:2024hix}. The original spectrum is rescaled by setting $\Delta_T=3$ and $\Delta_G=2.5$ for integer spin and half-integer spin sector respectively. 
    }
	\begin{tabular}{c|ccccc|ccc|cc}
		\hline
		\hline 
		Operator & $\sigma$ & $\epsilon$ & $\sigma'$ & $\epsilon'$ & $\epsilon''$ & $\psi$ & $\psi'$ & $G_\mu$ & $T_{\mu\nu}$ & $T'$ \\
		$L$ & 0 & 0 & 0 & 0 & 0 & 1/2 & 1/2 & 3/2 & 2 & 2 \\ \hline 
		$\Delta_{\text{CB}}$ & 0.584 & 1.584 & 2.887 & 3.887 & 4.38 & 1.084 & 3.387 & 2.5 & 3.0  & 3.4    \\
		$\Delta_{\text{FS}}$ & 0.591 & 1.516 & 2.868 & 3.929 & 4.031 & 1.109 & 3.389 & 2.5* & 3.0* & 3.449    \\ 
		Error & 1.21$\%$ & 4.27$\%$ & 0.67$\%$ & 1.09$\%$ & 7.97$\%$ & 2.31$\%$ & 0.07$\%$ & - & -  & 1.44$\%$ \\ 
		\hline
	\end{tabular}
	\label{tab:numerical2}
\end{table*}

\end{document}